\documentclass{sig-alternate-2013}
\usepackage{amsmath}
\usepackage{graphicx}
\usepackage{amssymb}
\usepackage{url}
\makeatletter
\g@addto@macro{\UrlBreaks}{\UrlOrds}
\makeatother
\usepackage{mathrsfs}
\DeclareMathAlphabet{\mathpzc}{OT1}{pzc}{m}{it}
\usepackage{subfigure}
\usepackage{stmaryrd}
\usepackage{algorithm}
\usepackage[noend]{algorithmic}

\permission{Copyright is held by the International World Wide Web Conference Committee (IW3C2). IW3C2 reserves the right to provide a hyperlink to the author's site if the Material is used in electronic media.}
\conferenceinfo{WWW 2016,}{April 11--15, 2016, Montr\'eal, Qu\'ebec, Canada.} 
\copyrightetc{ACM \the\acmcopyr}
\crdata{978-1-4503-4143-1/16/04. \\
http://dx.doi.org/10.1145/2872427.2882996}
\begin{document}
\title{Mining Online Social Data for Detecting Social Network Mental Disorders}
\author{Hong-Han Shuai$^1$, Chih-Ya Shen$^1$, De-Nian Yang$^1$, Yi-Feng Lan$^2$, \\ 
	Wang-Chien Lee$^3$, Philip S. Yu$^{4,5}$, Ming-Syan Chen$^6$ \\
	%EndAName
	\affaddr{$^1$Academia Sinica, Taiwan} \\
	\affaddr{$^2$Tamkang University, Taiwan} \\
	\affaddr{$^3$The Pennsylvania State University, USA} \\ 
	\affaddr{$^4$University of Illinois at Chicago, USA} \\
	\affaddr{$^5$Tsinghua University, China} \\
	\affaddr{$^6$National Taiwan University, Taiwan} \\
	\affaddr{hhshuai@citi.sinica.edu.tw, chihya@citi.sinica.edu.tw, dnyang@iis.sinica.edu.tw} \\
	\affaddr{carolyflan@gmail.com, wlee@cse.psu.edu, psyu@uic.edu, mschen@cc.ee.ntu.edu.tw}
}
\maketitle

\begin{abstract}
An increasing number of social network mental disorders (SNMDs), such as Cyber-Relationship Addiction, Information Overload, and Net Compulsion, have been recently noted. Symptoms of these mental disorders are usually observed passively today, resulting in delayed clinical intervention. In this paper, we argue that mining online social behavior provides an opportunity to actively identify SNMDs at an early stage. It is challenging to detect SNMDs because the mental factors considered in standard diagnostic criteria (questionnaire) cannot be observed from online social activity logs. Our approach, new and innovative to the practice of SNMD detection, does not rely on self-revealing of those mental factors via questionnaires. Instead, we propose a machine learning framework, namely, \textit{Social Network Mental Disorder Detection (SNMDD)}, that exploits features extracted from social network data to accurately identify potential cases of SNMDs. We also exploit multi-source learning in SNMDD and propose a new SNMD-based Tensor Model (STM) to improve the performance. Our framework is evaluated via a user study with 3126 online social network users. We conduct a feature analysis, and also apply SNMDD on large-scale datasets and analyze the characteristics of the three SNMD types. The results show that SNMDD is promising for identifying online social network users with potential SNMDs.
\end{abstract}
%\begin{IEEEkeywords}
%Supervised Machine Learning; Social Network Analysis; Mental Health.
%\end{IEEEkeywords}
%\BASELINESKIP=12PT
 
%\terms{Algorithms, Experimentation}
\keywords{Online social network, mental disorder detection, feature extraction, tensor factorization}

\section{Introduction}
``\textit{As we expect more from technology, do we expect less from each other?}'' asked Sherry Turkle, the Abby Rockefeller Mauz\'e professor of the Social Studies of Science and Technology in MIT.\footnote{\url{http://www.ted.com/talks/sherry_turkle_alone_together}} With the explosive growth in popularity of social networking and messaging apps, online social networks (OSNs) have become a part of many people's daily lives. While OSNs seemingly expand their users' capability in increasing social contacts, they may actually decrease the face-to-face interpersonal interactions in the real world. Studies show that some people's behavior is bolder on the OSNs because they can put a mask when communicating with others there, i.e., hide who they really are. However, do those OSN users still know how to connect with others when the masks are off? Lying between receiving positive attention from OSNs and face-to-face interactions may be a great gulf in the real life.

Most research on social network mining focuses on discovering the treasure of knowledge behind the data for improving people's life. In contrast, much less attention has been drawn to remedy the problems incurred from various social network applications. Indeed, some \emph{social network mental disorders (SNMDs)}, such as Information
Overload and Net Compulsion~\cite{YoungType98}, have been recently noted.\footnote{\url{http://phys.org/news/2015-09-social-media-impacts-mental-well-being.html}} For example, studies suggest that 1 in 8 Americans suffers from problematic Internet use.\footnote{\url{http://netaddiction.com/faqs/}} Moreover, as reported by the BBC News, 25\% of the population in Korea are estimated to suffer from SNMDs.\footnote{\url{http://www.bbc.com/news/world-asia-33130567}} Due to the epidemic scale of these phenomena, new terms such as Phubbing (Phone Snubbing) and Nomophobia (No Mobile Phone Phobia) have been created to describe those who cannot stop using mobile social networking apps. Moreover, leading journals in mental health, such as American Journal of Psychiatry \cite{b08}, have reported that the SNMDs may incur excessive use, depression, social withdrawal, and a range of negative repercussions. Indeed, these symptoms are important components of diagnostic criteria for detecting SNMDs~\cite{Kimberly98}:
1) \textit{excessive use} of social networking and messaging apps -- usually associated with a loss of
the sense of time or a neglect of basic drives; 2) \textit{withdrawal} --
including feelings of anger, tension, and/or depression when the computer/apps are inaccessible; 
3) \textit{tolerance} -- manifesting as the need for more usage; 
and 4) \textit{negative repercussions} -- including arguments, lying, social isolation, and fatigue~\cite{b08}. SNMDs are social-oriented and tend to happen to users who usually interact with others via online social media. Those with SNMDs usually lack offline interactions, and as a result seek cyber-relationships to compensate.

Today, identification of potential mental disorders often falls on the shoulders of supervisors (such as teachers, employers, or parents) who can observe the aforementioned symptoms better than others but only passively. As the facts that there are very few notable physical risk factors, the patients usually do not actively seek medical or psychological services to reduce these symptoms. Consequently, patients would look for clinical interventions with psychiatrists and medical treatments only when their conditions become very serious. However, a recent study \cite{Lin14} shows a strong correlation between suicidal attempt and SNMDs for students. In this research, 9510 adolescent students aged from 12 to 18 years old are tested using a personality inventory and Internet addiction inventory. The findings indicate that adolescents suffering from social network addictions have a much higher risk of suicidal ideation than non-addictive users. The research also reveals that social network addiction may deteriorate emotional status, causing higher hostility, depressive mood and compulsive behavior. Most importantly, the delay of early intervention may lead to mental illness and thus can seriously damage an individual's social functioning. In short, it is desirable to actively detect potential SNMD users on OSNs at an early stage. 

Although previous work in Psychology has identified several crucial mental factors related to SNMDs as standard diagnostic criteria for detecting SNMDs, they are mostly assessed via survey questionnaires by design. To detect potential SNMD cases of OSN users, extracting these factors to assess the mental states of users is very challenging. For example, the extent of loneliness and the effect of disinhibition of OSN users are not easily observable.\footnote{The online disinhibition effect is a loosening (or complete abandonment) of social restrictions and inhibitions that would otherwise be present in normal face-to-face interaction during interactions with others on the Internet.}

% While using the online duration for evaluating SNMDs looks intuitive, it is a poor indicator as reported in \cite{Lai13}.

%In this paper, we argue that mining online social behavior has a potential to \textit{actively identify} SNMDs at an early %stage.
% In contrast to the conventional approaches where the supervisors and psychiatrists observe the behavior of a patient %or have interventions with a patient passively, 
%the detailed online interactions and information access of users logged by social network applications can be effectively %exploited to provide an early alert. In other words, mining 
There is a need for developing new approaches for detecting SNMD cases of OSN users. We argue that mining social network data of individuals, as a complementary alternative to the conventional psychological approach, provides an excellent opportunity to {\em actively identify} those cases at an early stage. 
%To the best of our knowledge, there has been no data mining research or software today addressing this important %need, especially when an increasing number of people, including children and students, are becoming more inclined to %interact with each other via OSNs, instead of offline face-to-face meetings.
In this paper, we 
%make the first attempt to automatically identify potential online users with SNMDs and 
develop a machine learning framework for detecting SNMDs, namely \textit{Social Network Mental Disorder Detection (SNMDD)}. We formulate the task as a semi-supervised classification problem to detect three types of SNMDs~\cite{YoungType98}, including i) Cyber-Relationship Addiction, which shows addictive behavior for building online relationships; ii) Net Compulsion, which shows compulsive behavior for online social gaming or gambling; and iii) Information Overload, which is related to uncontrollable surfing. By exploiting machine learning techniques with the ground truth obtained via the current diagnostic practice in Psychology \cite{YoungType98}, we extract and analyze several features of different categories from OSNs, including parasocial relationships, online and offline interaction ratio, social capital, disinhibition, self-disclosure, and bursting temporal behavior. These features capture important factors or serve as proxies for SNMD detection. For example, parasocial relationship represents an asymmetric interpersonal relationship where one party cares more about the other, but the other does not. This asymmetric relationship is related to loneliness, one of the primary mental factors for the users with SNMDs to access online social media excessively~\cite{Baek13}. Therefore, we extract the ratio of the number of actions a user takes to friends and the number of actions friends take to the user as a feature. In this paper, the extracted features are carefully examined through user study.

Users may behave differently on different OSNs, resulting in inaccurate SNMD detection. When the data on different OSNs of a user are available, the accuracy of SNMDD is expected to improve by effectively integrating information from multiple sources for model training. A na\"{i}ve solution is to concatenate the features from different networks into a feature vector. However, simply increasing the number of features suffers from the curse of dimensionality. Accordingly, we propose an \textit{SNMD-based Tensor Model (STM)} to deal with this multi-source learning problem in SNMDD. There are two advantages of our approach: i) the novel \textit{STM} incorporates the SNMD characteristics into the tensor model; and ii) tensor factorization can capture the structure, latent factors, and correlation of features to derive a full portrait of user behavior.

The contributions of this paper are summarized as follows.

\begin{itemize}
\item Today online SNMDs are usually treated at a late stage. To address this issue, we propose an approach, new to the current practice of SNMD detection, by mining data logs of OSN users to actively identify potential SNMD cases early. 

\item We develop a machine learning framework for detecting SNMDs, namely \textit{Social Network Mental Disorder Detection (SNMDD)}. Moreover, we design and analyze many features from OSNs, such as disinhibition, parasociality, self-disclosure, etc., which serve as important factors or proxies for identifying SNMDs. The proposed framework can be deployed as a software program to provide an early alert for potential patients and their advisors.

\item We study the \emph{multi-source learning} problem for SNMD detection. By leveraging tensor algebra and considering the SNMD characteristics into the tensor model, we propose \textit{STM} to better extract the latent factors from different sources, thus improving the accuracy.

\item We conduct a user study with 3126 users to evaluate the effectiveness of the proposed SNMDD framework. To the best of our knowledge, this is the first dataset crawled online for SNMD detection. Also, we apply SNMDD on large-scale real datasets and perform a social network analysis on the detected cases. The result reveals interesting insights on the network structures in SNMD types, which can be of interest to social scientists and psychologists.
\end{itemize}

The rest of this paper is organized as follows. Section \ref{sec:02_relatedwork} surveys the related work. Section \ref{sec:03SNMD} presents \emph{SNMDD}, focusing on feature extraction and the proposed \textit{STM} for multi-source learning. Section \ref{sec5:exp} reports a user study, various analyses, and the experimental results. Section \ref{conclu} concludes this paper.

\section{Related Work}
\label{sec:02_relatedwork}
Recent research in Psychology and Sociology reports a number of mental factors related to social network mental disorders. Specifically, Barbera et al. find that young people with narcissistic tendencies are particularly vulnerable to addiction with OSNs \cite{Barbera09}. Moreover, Chak et al. show that the tendency of addiction to the Internet is positively related to the shyness, and negatively related to the faithfulness \cite{Chak04}.
%Additionally, several mental factors can effectively predict the intention to use OSNs as well as their actual usage, e.g., playfulness, trust in the site, and perceived ease of use \cite{Sledgianowski09}.
Note that most of these features cannot be directly observed in OSNs. Thus, in our study, we investigate new features in social network data for SNMD detection. On the other hand, statistical analysis on the prevalence and extent of mobile email addictions has been studied \cite{cacm12}, and a quantitative analysis of the reasons to leave Facebook with over 400 questionnaires was performed to improve human-computer interactions on social media \cite{Baumer13}. Our research in this paper is uniquely different from these prior works since we adopt a data mining approach to detect SNMDs by exploiting various discriminative features to capture mental factors. Moreover, we propose a tensor model to efficiently integrate heterogeneous data from different OSNs and incorporate the domain knowledge, i.e., the SNMD characteristics such as the users with Cyber-Relationship Addiction tend to have friends with Cyber-Relationship Addiction.

A recent line of studies proposes to detect cyberbullying, i.e., harassing other OSN users in a deliberate manner. Dinakar et al. \cite{Dinakar11} propose a machine learning approach and extract features from content sentiment and context information to detect textual cyberbullying. The detection of SNMDs is more difficult than that of textual cyberbullying because i) it is possible to focus on certain keywords to detect cyberbullying behavior, and ii) users cannot hide the cyberbullying behavior (it is not cyberbullying if users can hide the behavior), whereas users with Net Compulsion may hide their logs. 

Our framework is built upon support vector machine, which has been widely used to analyze OSNs in many areas \cite{SVM1,SVM3}, such as business, transportation, and anomaly intrusion detection. In addition, we present a new tensor model that not only incorporates the domain knowledge but also well estimates the missing data and avoids noises to properly handle multi-source data. Crammer et al. propose a PAC-style model for multi-source learning and provide a theory of sampling for learning models \cite{Crammer08}. However, the SNMD data from different OSNs may be incomplete due to the heterogeneity. For example, the profiles of users may be empty due to the privacy issue, different functions on different OSNs (e.g., game, check-in, event), etc. We propose a novel tensor-based approach to address the issues of using heterogeneous data and incorporate domain knowledge in SNMD detection.

%Kumar et al. use the idea of co-training for solving the multi-view clustering problem via spectral projection \cite{Kumar11}.

%%%Section 3
\section{Social Network Mental Disorder Detection}
\label{sec:03SNMD}
%Social media has attracted an enormous number of users. The participatory
%nature of social media makes people highly conducive to share information,
%express their feelings, and accumulate their social capital. As a result,
%While logged data of online social behavior provide an opportunity to identify SNMDs for early treatment intervention, it is still
%difficult for mental health professionals to directly interpret online social data.
In this paper, we aim to explore data mining techniques to detect three types of SNMDs~\cite{YoungType98}: 
1) \textit{Cyber-Relationship (CR) Addiction}, which includes addiction to
social networking, checking and messaging to the point where social
relationships to virtual and online friends become more important than
real-life ones with friends and families; 2) \textit{Net Compulsion (NC)}, which
includes compulsive online social gaming or gambling, often resulting in
financial and job-related problems; and 3) \textit{Information Overload (IO)}, which
includes addictive surfing of user status and news feeds, leading to lower
work productivity and fewer social interactions with families and friends
offline.

\begin{figure}[t]
\centering
\includegraphics[scale=0.3] {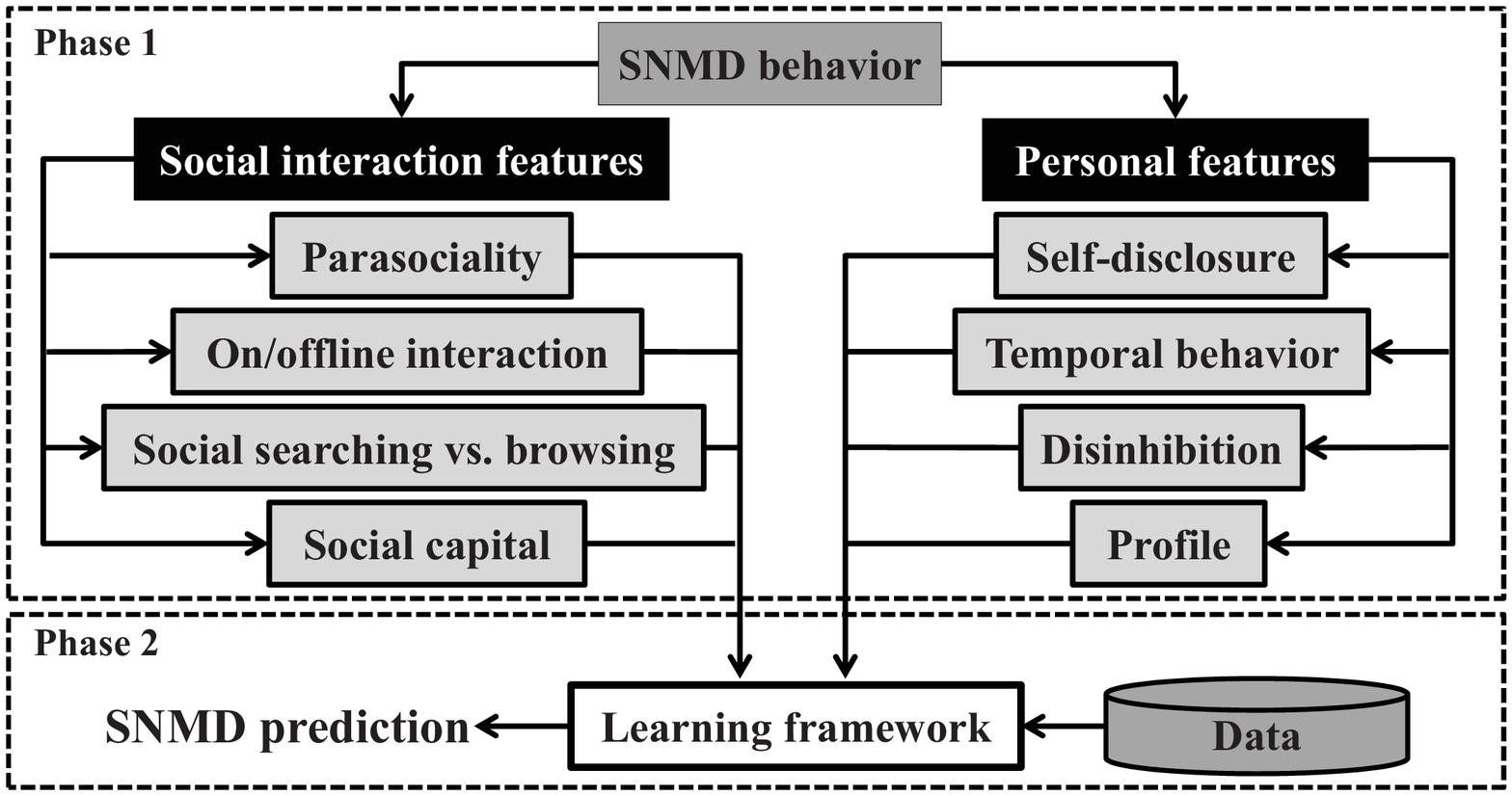}
%\vspace{-10pt}
\vspace{-2mm}
\caption{The SNMDD framework.}
\label{figframework}
\vspace{-3mm}
\end{figure}

Accordingly, we formulate the detection of SNMD cases as a classification problem. We detect each type of SNMDs with a binary SVM. In this study, we propose a two-phase framework, called \textit{Social Network Mental Disorder Detection (SNMDD)}, as shown in Figure \ref{figframework}. The first phase extracts various discriminative features of users, while the second phase presents a new SNMD-based tensor model to derive latent factors for training and use of classifiers built upon Transductive SVM (TSVM) \cite{TSVM10}. Two key challenges exist in design of SNMDD: i) we are not able to directly extract mental factors like what have been done via questionnaires in conventional SNMD detection process for psychiatrists and thus need new features for learning the classification models;\footnote{Additional issues in feature extraction will be detailed later.} ii) we aim to exploit user data logs from multiple OSNs and thus need new techniques for integrating multi-source data based on SNMD characteristics. We address these two challenges in Sections \ref{sec:031_feature} and \ref{sec:35tensordecomposition}, respectively.

\subsection{Feature Extraction}
\label{sec:031_feature}

We first focus on extracting discriminative and informative features for design of SNMDD. This task is nontrivial for the
following three reasons.

\noindent \textbf{1. Lack of mental features.} Psychological studies have shown that
many mental factors are related to SNMDs, e.g., low self-esteem \cite{Kimberly98}, loneliness \cite{Leung04}. Thus, questionnaires are designed to reveal those factors for SNMD detection.
Some parts of Psychology questionnaire for SNMDs are based on subjective comparison of mental states in online and offline status, which cannot be observed from OSN logs. For example:

\vspace{1.7mm}
\noindent \textbf{Q1.} How often do you feel depressed, moody, or nervous when you are off-line, which goes away once you are back online? \\
\noindent \textbf{Q2.} How often do you prefer the excitement of the Internet to intimacy with your partner?
\vspace{1.7mm}

Consider Q1. The feel of depression and nervousness offline can not be observed online. To tackle this problem, we have to leverage the knowledge from Psychology, such as withdrawal or relapse patterns, and exploit some proxy features extracted from online social activity logs to approximate them. For Q2, the preference of excitement of the Internet to intimacy with users' partners is important questions for SNMD detection. As it is difficult to directly observe these factors from data collected from OSNs, psychiatrists are not able to directly assess the mental states of OSN users under the context of online SNMD detection.

\noindent \textbf{2. Heavy users vs. addictive users.} To detect SNMDs, an intuitive idea is to simply extract the usage (time) of a user as a feature for training SNMDD. However, this feature is not sufficient because i) the status of a user may be shown as ``online'' if she does not log out or close the social network applications on mobile phones, and ii) heavy users and addictive users all stay online for a long period, but heavy users do not show symptoms of anxiety or depression when they are not using social apps. How to distinguish them by extracting discriminative features is critical.

\noindent \textbf{3. Multi-source learning with the SNMD characteristics.} As we intend to exploit user data from different OSNs in SNMDD, how to extract complementary features to draw a full portrait of users while considering the SNMD characteristics into the tensor model is a challenging problem.

To address the first two challenges, we identify a number of effective features as proxies to capture the mental states of users, e.g., self-esteem \cite{Kimberly98} and loneliness \cite{Leung04}.\footnote{The third challenge is addressed in Section \ref{sec:35tensordecomposition}.} The goal is to distinguish users with SNMDs from normal users. Two types of features are extracted to capture the social interaction behavior and personal profile of a user.

\subsubsection{Social Interaction Features}
\label{sec:0312_social}

We first extract a number of \emph{social interaction features} to capture the user behavior on social media.

\noindent\textbf{Parasocial relationship (PR).} Research shows that the mental factor of loneliness is one of the primary reasons why the users with SNMDs excessively access online social media \cite{Baek13}. As the loneliness of an OSN user is hard to measure, we exploit the parasocial relationship, an asymmetric interpersonal relationship between two people where one party cares more about the other but the other does not, to capture loneliness (as studies show that they are correlated~\cite{Cacioppo09}). The feature of parasocial relationship is represented as $%
|a_{out}| / |a_{in}|$, where $|a_{out}|$ and $|a_{in}|$ denote the number of actions a user takes to friends and the number of actions friends take to the user, respectively.\footnote{The actions include like, comment, and post in our work.} As the ratio increases, the extent of parasocial relationship also grows.

\noindent\textbf{Online and offline interaction ratio (ONOFF).} As observed by
mental health professionals, people who indulge themselves in OSNs tend to snub their friends in real life \cite{Chak04}. Therefore, the number of online
interactions is inclined to significantly exceed their interactions offline.
We extract the number of check-in logs with friends and the number
of ``going'' events as an indicator of the number of offline activities to estimate the online ($|a_{on}|$)/offline ($|a_{off}|$) interaction  ratio. Although the number of offline events observed from online is smaller than the actual number, the ratio is relative and is a good indicator (as pointed out in \cite{CheckinD}), because the frequent check-in records of a user imply that the user is active in offline activities, which is an indicator of non-SNMD.

\noindent\textbf{Social capital (SC).} Two types of friendship ties are usually involved in the theory of \emph{social capital} \cite{Steinfield12}: i) Bond strengthening
(strong-tie), which represents the use of OSNs to strengthen
the relationships; and ii) Information seeking (weak-tie), which
corresponds to the use of social media to find valuable information. The first type usually creates more interactions in order to increase the social tightness and is related to Cyber-Relationship (CR) Addiction, while the second type concentrates more on finding and reading the information and is thus related to Information Overload (IO) \cite{socialcap01}. Therefore, the ratio between the number of strong ties ($n_{strong}$) and weak ties ($n_{weak}$) could be used for differentiating the CR and IO types. Moreover, since the number of strong ties is much smaller than that of weak ties \cite{Arnaboldi13}, and the number of friends that CR users frequently interact with is less than that of IO users, we exploit the ratio between the number of friends the user interacts online (likes, comments, and posts) and the total number of the user's friends as proxy features to differentiate the CR and IO types.

\noindent\textbf{Social searching vs. browsing (SSB).} The human appetitive system is in charge of the addictive behavior. A recent study has shown that \emph{social searching} (actively reading news feeds from friends' walls) creates more pleasure than \emph{social browsing} (passively reading personal news feeds)~\cite{Wise10}. This finding indicates that goal-directed activities of social searching are more likely 
to activate the appetitive system of a person as drug rewards do, and it is more related to SNMDs because the appetitive system is responsible for finding things in the environment that promote species survival (i.e., food, sexual mates) and thus is inclined to form addictive behavior after several rewards. While users with SMNDs perform social searching more frequently than non-SNMDs, it is not easy to distinguish these two behavior on social media. Let $n_{i}$ denote the total number of the $i$-th action for posts among friends. For example, if a user is the second one among her friends who click ``likes'' on a post, the $n_{2}$ increases $1$ for the user. As most social media provide friends' comments and ``likes'' in the form of news feeds to users, we consider the number of likes/comments on news feeds from friends as social browsing ($\sum _{i=2}^{\infty} n_{i}$). On the other hand, if users take an initiative to search for someone's profile and like/comment on it, we consider this as a social searching (i.e., the number of likes/comments on others' news feeds that are not liked/commented by his friend before ($n_1$)). Therefore, we use $\frac{n_{1}}{\sum _{i=2}^ \infty n_{i}}$ as a feature. The social searching features are related to CR because CR users tend to find social supports, whereas social browsing is more related to IO. Compared with social capital, SSB focuses on different behavior in reading news feeds, rather than the different types of friend ties.

\subsubsection{Personal features}
\label{sec:0311_personal}

In the following, we present the features extracted from the personal profiles of OSN users.

\noindent\textbf{Self-disclosure based features (SD).} Researchers from Harvard University
point out that a person's \emph{self-disclosure communications} (i.e.,
describing the personal feeling) stimulate the brain's
pleasure center \cite{Tamir12}, similar to sex and food. However, to conduct a 
sentiment analysis on the contents associated with a user is very
complicated and computationally expensive. Inspired by emotional signal detection, which finds that when the users use emoticons, they are effectively expressing an emotional state \cite{LiuEmoticon12}, we retrieve and exploit the numbers of emoticons, stickers, and selfies in each post as the features for self-disclosure \cite{Seidman13}.

%In summary, we extracted the following four features. i) \textit{Number of used emoticons.} Users in social media usually express their emotional states with visual cues, i.e., emoticons. Given that expressions of users' emotional states are highly related to emoticons, we regard the average number of emoticons per post as a feature for measuring emotional expressions as self-disclosure. ii) \textit{Number of used stickers (smiley).} Stickers are similar to emoticons but are usually larger and associated with different characters. We adopted user stickers per comment as an alternative feature for self-disclosure because  a) stickers are more commonly-used than emoticons nowadays in many social media, e.g., LINE, Facebook, and b) it is practical to extract stickers from comments with the label of \textit{sticker$\_$id}. iii) \textit{Selfies.} Since selfies and posts represent the behavior of self-disclosure, we took the number of selfies, number of likes on selfies, and number of comments on selfies as features. iv) \textit{Ratio between like and comment.} "Likes" in social media highlight the content of which users are fond of, though for some users, it only represents a ``read'' behavior to show their concern or awareness without writing a comment. Therefore, if a user clicks ``likes'' on a lot of news feeds without commenting on any of them, the ratio becomes large, indicating that the user may be browsing information online rather than strengthening the bonds.

\noindent\textbf{Temporal behavior features (TEMP).} \emph{Relapse} is the state that a person is inclined to quickly revert back to the excessive usage of social media after an abstinence period, while \emph{tolerance} is the state that the
time spent by a person with SNMDs tends to increase due to the mood
modification effect.\footnote{A patient may need to spend more time on social media to reach the happiness/excitement than before.} It is worth noting that the above two mental states have been exploited to evaluate clinical addictions \cite{Leung04}. We aim to use them to distinguish \emph{heavy users} and \emph{addictive users} because heavy users do not suffer from relapse and tolerance in use of OSNs. An issue arising here is how to assess relapse and tolerance quantitatively.

It is observed that the use of social media by an SNMD patient is usually in the form of \textit{intermittent bursts}~\cite{Kimberly98}. Therefore, given a stream of a user's activities on an OSN, e.g., ``likes'', ``comments'', ``posts'', we exploit Kleinberg's burst detection algorithm \cite{Kleinberg02}, which is based on an infinite Markov model, to detect periods of the user's activities as bursty and non-bursty periods. The bursty period refers to a period during which the activities significantly increase. A bursty period is modeled as a bursty state $q_1$ in the Markov model, while a non-bursty period is correspondingly modeled as a normal state $q_0$. The burst detection algorithm finds a state transition sequence $\mathbf{q}$ for each user to divide the corresponding log (stream of activities) into bursty and non-bursty periods. Specifically, let $\mathbf{x}=(x_1,x_2,\cdots,x_n)$ denote a sequence of $n$ time intervals between $n+1$ consecutive activities, with the intervals distributed according to a density function, such as $f_{i_t}(x_t)=\alpha_{i_t}e^{-\alpha_{i_t}x_t}$, where $\alpha_{i_t}$ is either $\alpha _0$ or $\alpha _1$, and $\alpha_0$ and $\alpha_1$ are parameters that correspond to normal and burst states, respectively, $\alpha_1>\alpha_0$. A time interval $x_t$ is in a burst state $q_1$ if $f_0(x_t)$ < $f_1(x_t)$. Otherwise, it is in a normal state $q_0$. However, simply deciding the state sequence $\mathbf{q}$ based on this criteria results in numerous small periods. Therefore,
a cost $\tau(q_i,q_j)$ is associated with a state transition from $q_i$ to $q_j$ to filter out noises and to ensure that each bursty or non-bursty period is sufficiently long. Therefore, the remaining issue is to find an optimal state-transition sequence $\mathbf{q}$ to minimize the following cost function \cite{Kleinberg02},

\begin{equation*}
c(\mathbf{q}|\mathbf{x})=\sum_{t=1}^{n-1}\tau(q_i,q_{i+1})+\sum_{t=1}^n(-\ln f_{i_t}(x_t)),
\end{equation*}
where $\tau(q_i,q_{i+1})=0$ if the state $q_i$ and $q_{i+1}$ is the same. $\tau(q_i,q_{i+1})$ is $\gamma \ln n$ otherwise, where $\gamma$ is an algorithm parameter larger than $0$. Notice that the state sequence that minimizes the cost depends on 1) how easy it is to jump from one state to another and 2) how well it is to comply to the rates of arrivals. After identifying the bursts, we measure their intensity (the number of activities within a burst) and length (the time period of a burst) as the proxy features for \emph{relapse} and \emph{tolerance}, respectively. The $\langle$average, median,
standard deviation, maximum, minimum$\rangle$ of both burst intensity and burst length are included in our feature set, because they capture the characteristic of bursts. For instance, the standard deviation of the burst length for SNMD patients is usually larger than that for heavy users since heavy users constantly use OSNs while users with SNMDs increase the usage time due to tolerance.

\noindent\textbf{Usage time (UT).} In addition to the above, two features regarding usage of an OSN are also adopted. i) \textit{Duration.} The duration that a
user spends on social media a day is estimated by consecutive activity logs.\footnote{If the difference of timestamps of two logs is smaller than a few minutes, the user is regarded as online during this period \cite{Klinkner08}.} ii)
\textit{Number of online states.} The number of online
states during a day is also important. It was widely believed that a person spending a lot of time on OSNs usually belongs to CR or IO. However, a recent study points out that the usage time is only moderately correlated to CR and IO~\cite{Lai13}. Indeed, both heavy users and the users with SNMDs tend to stay online for lengthy time periods, but heavy users do not feel anxious and depressed when they are not using social apps. With the bursts detected earlier, we aim to use the relapse and tolerance to distinguish users with SNMDs from heavy users. For example, given a heavy user spends the same amount of time online as a user with SNMDs does, the standard deviation of burst length for the heavy user is expected to be smaller than that of the user with SNMDs since the heavy user explores OSNs more regularly and does not suffer from the tolerance of increasing usage.

\noindent\textbf{Disinhibition based features (DIS).} The mental factor of disinhibition is also one of the primary reasons the users excessively access online social media \cite{Baek13}. When surfing online, some people tend to act out more frequently or intensely than they act offline due to the dissociative anonymity, asynchronicity, and solipsistic introjection, which is called the \textit{online disinhibition effect} \cite{Suler04}. When user identities can be anonymous or the conversation is not face-to-face (e.g., Whisper, SnapChat), offline-shy users are more inclined to addict to cyberspace relationships due to disinhibition. Although we know which OSNs are anonymous and thus expect a stronger disinhibition effect on them, how to detect the users experiencing the disinhibition effect from non-anonymous OSNs is challenging. As reported in \cite{Wang14}, the average clustering coefficient (CC) on anonymous OSNs is smaller than that on non-anonymous ones, e.g., 0.033 on Whisper but 0.059 on Facebook. Inspired by this observation, we use CC as a proxy for disinhibition on non-anonymous websites since a user with disinhibition tends to have a small CC. The disinhibition effect is more related to CR and NC because they both show stronger intensity of usage under the anonymity. Also, the clustering coefficient may potentially be effective for detecting CR and NC because it depends on the links among friends which cannot be hidden by users.

%Also, in order to achieve disinhibition, users usually create multiple accounts for hiding personal information from real friends. For example, users may create multiple accounts and connect them to a real account in order to receive tickets to complete a stage in Candy Crush. Another example is that a user may create new accounts to make friends with strangers so that their friends will not know. To extract the number of multiple accounts as a feature, one basic approach is to identify the same users across different social networks by the state-of-the-art \cite{ZL13,JKJ13}. Moreover, since the accounts for completing games may have only one friend (the real account) or so, we extracted the number of a user's friends who has only one friend as a feature.

\noindent\textbf{Profile features (PROF).} We also extract some demographic features commonly adopted in questionnaires from user profiles, such as \textit{age} and \emph{gender}. A study~\cite{Leung04} has shown that the age when a user logs in Facebook the first time is correlated to the intensity of SNMDs. For children growing up with the Internet, this is a key position in their lives. Empirical studies~\cite{Leung04} also suggest that gender difference results in varying degrees of SNMDs because the goals of using OSNs are different for different genders, e.g., females are more inclined to use online communication whereas males are inclined to follow news and play online games. Also, the number of game posts is extracted.

\noindent \textbf{Feature Effectiveness.} It is worth noting that each individual feature cannot precisely classify all cases, as research shows that exceptions may occur. Therefore, it is necessary to exploit multiple features to effectively remove exceptions. For example, the social searching vs. browsing (SSB) feature may be affected by the gender difference of using OSNs, i.e., males are more inclined to social searching than females \cite{Makashvili13}. Moreover, when using the number of game logs as an indicator for NC, a special case happens when the users with NC attempt to hide the game logs by not allowing games to publish news feeds. However, we are able to address the above issue by exploiting other features, e.g., CC for disinhibition effect, which is dependent on the user's friends so that the user cannot change it. Therefore, the special cases on existing individual features can be overcome by fusing multiple features together.

\subsection{Multi-Source Semi-Supervised Learning}
\label{sec:35tensordecomposition} Many users are inclined to use different
OSNs, and it is expected that data logs of these OSNs could provide enriched and
complementary information about the user behavior. Thus, we aim to explore
multiple data sources (i.e., OSNs) in SNMDD, in order to derive a more
complete portrait of users' behavior and effectively deal with the data sparsity
problem. To exploit multi-source learning in
SNMDD, one simple way is to directly concatenate the features of each person
derived from different OSNs as a huge vector. However, the above approach
tends to miss the correlation of a feature in different OSNs and introduce
interference. Thus, we explore tensor techniques which have been used increasingly to model multiple data sources because a tensor can
naturally represent multi-source data. We aim to employ tensor decomposition to
extract common latent factors from different sources and objects. 

In this paper, given $D$ SNMD features of $N$ users extracted from $M$ OSN sources, we construct a three-mode tensor $\mathcal{T}\in \mathbb{R%
}^{N\times D\times M}$ and then conduct Tucker decomposition, a renowned tensor decomposition technique, on $\mathcal{T}$ to extract a latent feature matrix $%
\mathbf{U}$, which presents \emph{the latent features of each person summarized
from all OSNs}. We aim to feed these latent features for SNMD detection. Matrix $\mathbf{U}$ effectively estimates a deficit
feature (e.g., a missing feature value unavailable due to privacy setting) of an OSN
from the corresponding feature of other OSNs, together with the features of
other users with similar behavior. Based on Tucker decomposition on $%
\mathcal{T}$, we present a new \textit{SNMD-based Tensor Model (STM)}, which enables $\mathbf{U}$ to incorporate important characteristics of
SNMDs, such as the correlation of the same SNMD sharing among close friends.\footnote{Note that $D$ does not capture the social correlations among friends.}
Finally, equipped
with the new tensor model, we conduct semi-supervised learning to classify
each user by exploiting Transductive Support Vector Machines
(TSVM).

We first summarize the notations introduced in this section. Here scalars
are denoted by lowercase letters, e.g., $u$, while vectors are denoted by
boldface lowercase letters, e.g., $\mathbf{u}$. Matrices are represented by
boldface capital letters, e.g., $\mathbf{U}$, and tensors are denoted by
calligraphic letters, e.g., $\mathcal{T}$. Each element $%
(i,j,k)$ of a three-mode tensor $\mathcal{T}$ is denoted by $t_{ijk}$%
, whereas the $i$-th row and the $j$-th column of a two-dimensional matrix $\mathbf{U}$ are
respectively denoted by $\mathbf{u}_{i:}$ and $\mathbf{u}_{:j}$.
\sloppy
Specifically, Tucker decomposition \cite{TuckerModel09} of a tensor $\mathcal{T}\in 
\mathbb{R}^{N\times D\times M}$ is defined as: 
\begin{equation}
\mathcal{T}=\mathcal{C}\times _{1}\mathbf{U}\times _{2}\mathbf{V}\times _{3}%
\mathbf{W,}
\label{TuckerDecom}
\end{equation}%
where $\mathbf{U}\in \mathbb{R}^{N\times R}$, $\mathbf{V}\in \mathbb{R}%
^{D\times S}$ and $\mathbf{W}\in \mathbb{R}^{M\times T}$ are latent
matrices. In this paper, the matrix of users' latent features $\mathbf{U}$ plays a crucial role. 

In Tucker decomposition, $R$, $S$, and $T$ are parameters to be set according to different criteria \cite{TuckerModel09}. The $1$%
-mode product of $\mathcal{C}\in \mathbb{R}^{R\times S\times T}$ and $%
\mathbf{U}\in \mathbb{R}^{N\times R}$, denoted by $\mathcal{C}\times _{1}%
\mathbf{U}$, is a matrix with size $N\times S\times T$, where each element $(%
\mathcal{C}\times _{1}\mathbf{U)}_{nst}=$ $\sum_{r=1}^{R}c_{rst}u_{rn}$.
Given the input tensor matrix $\mathcal{T}$ that consists of the features of
all users from every OSN, Tucker decomposition derives $\mathcal{C}$, $%
\mathbf{U}$, $\mathbf{V}$, and $\mathbf{W}$ to meet the above equality on $%
\mathcal{T}_{ndm}$ for every $n$, $d$, and $m$, where $\mathcal{C}$ needs to
be diagonal, and $U$, $V$, and $W$ are required to be orthogonal \cite%
{TuckerModel09}. By regarding $\mathbf{u}_{i:}$ in $\mathbf{U}$\ as the
latent features of user $i$, we can efficiently integrate the information
from different networks for $i$.

Equipped with tensor decomposition on $\mathcal{T}$, we propose a new
\textit{SNMD-based Tensor Model (STM)} to minimize the following objective function $%
\mathcal{L}$, 
\begin{multline}
\mathcal{L}(\mathbf{U},\mathbf{V},\mathbf{W},\mathcal{C})=\frac{1}{2}\Vert 
\mathcal{T}-\mathcal{C}\times _{1}\mathbf{U}\times _{2}\mathbf{V}\times _{3}%
\mathbf{W}\Vert ^{2} \\
+\frac{\lambda _{1}}{2}tr(\mathbf{U}^{T}\mathbf{L}_{a}\mathbf{U})+\frac{%
\lambda _{2}}{2}\Vert \mathbf{U}\Vert ^{2}  \label{Eq:objfunction},
\end{multline}%
where $tr(\cdot )$ denotes the matrix traces, the Frobenius norm of a tensor 
$\mathcal{T}$ is defined as $\Vert \mathcal{T}\Vert =\sqrt{<\mathcal{T},%
\mathcal{T}>}$, and $\lambda _{1}$ and $\lambda _{2}$ are parameters
controlling the contribution of each part during the above collaborative
factorization. $\mathcal{L}$ first minimizes the decomposition error, i.e., $%
\Vert \mathcal{T}-\mathcal{C}\times _{1}\mathbf{U}\times _{2}\mathbf{V}%
\times _{3}\mathbf{W}\Vert ^{2}$, for $\mathcal{T}$. Note that Eq. (\ref{TuckerDecom}) does not always need to hold since other crucial goals are also incorporated in the model. For example,
the term that minimizes $\Vert \mathbf{U}\Vert ^{2}$ is to derive a more
concise latent feature matrix and avoid overfitting, where $\Vert \mathbf{V}%
\Vert ^{2}$, $\Vert \mathbf{W}\Vert ^{2}$, and $\Vert \mathcal{C}\Vert ^{2}$
are not necessary to be reduced since only the latent feature matrix $\Vert 
\mathbf{U}\Vert ^{2}$ will be employed in the semi-supervised learning later
in this section.

The \textit{STM} is different from the conventional tensor models in the second
term of Eq. (\ref{Eq:objfunction}), where important
characteristics of SNMDs are incorporated. For example, the probability of finding CR cases around a CR patient is higher than that around a non-CR user due to the loneliness propagation \cite{Cacioppo09}. That is, CR users usually feel lonely and are more likely to establish friendships in cyberspace with other users with similar behavior. Since the nearby nodes with a great quantity of interactions tend to be the same (either CR or non-CR), it is
envisaged that the distance of $\mathbf{u}_{i:}$ and $\mathbf{u}_{j:}$ will
be small if the edge weight $a_{i,j}$ of $e_{i,j}$ is sufficiently large. Therefore, a
regularization (smoothing) term $\frac{1}{2}tr(\mathbf{U}^{T}\mathbf{L}_{a}%
\mathbf{U})$ is included in the model to achieve the above goal,%
\begin{eqnarray*}
\frac{1}{2}tr(\mathbf{U}^{T}\mathbf{L}_{a}\mathbf{U}) &=&tr(\mathbf{U}^{T}(%
\mathbf{D}-\mathbf{A})\mathbf{U}) \\
&=&\frac{1}{2}\sum_{i,j}||\mathbf{u}_{i:}-\mathbf{u}_{j:}||_{2}^{2}a_{ij}\\
&=& \sum_{i,j}\mathbf{u}_{i:}a_{ij}\mathbf{u}_{i:}^{T}-\sum_{i,j}\mathbf{u}%
_{i:}a_{ij}\mathbf{u}_{j:}^{T} \\
&=&\sum_{i}\mathbf{u}_{i:}d_{ii}\mathbf{u}_{i:}^{T}-\sum_{i,j}\mathbf{u}%
_{i:}a_{ij}\mathbf{u}_{j:}^{T}.
\end{eqnarray*}%
Notice that $tr(\mathbf{U}^{T}\mathbf{L}_{a}\mathbf{U})$ decreases when the distance of latent factors for any two close friends is small. However, this term does not enforce the distance of features between every two friends to be small since the objective function $\mathcal{L}$
significantly increases in this case due to a larger decomposition error
appearing in the first term. The distance of $u_{i:}$ and $u_{j:}$ tend to be smaller if users $i$ and $j$ interact frequently, i.e., $a_{ij}$ is large. Therefore, the above model is able to incorporate
the characteristics of SNMD and alleviate the data sparsity problem, i.e., the users with fewer social activities may benefit from the auxiliary information of their friends with abundant features.

To properly find $\mathcal{L}(\mathbf{U},\mathbf{V},\mathbf{W},\mathcal{C})$
in \textit{STM}, let $\mathbf{A}$ denote the weighted
adjacency matrix of graph $G$.\footnote{%
The edge weight can be derived according to the number of interactions that
represents the proximity \cite{Chaoji}.} $\mathbf{L}_{a}=\mathbf{D}-\mathbf{Z}
$ is the Laplacian matrix of the weighted adjacency matrix $\mathbf{A}$,
where $D$ is a diagonal matrix with the entries $d_{ii}=\sum_{i}a_{ij}$. We
present a gradient-descent algorithm to iteratively improve each element
in the matrices according to the corresponding gradient, where the gradient
for each variable is derived as follows:
%U
\begin{multline*}
\nabla _{u_{i:}}\mathcal{L}=(\mathcal{C}\times _{1}\mathbf{u}_{i:}\times _{2}%
\mathbf{v}_{j:}\times _{3}\mathbf{w}_{k:}-t_{ijk})\mathcal{C}\times _{2}%
\mathbf{v}_{j:}\times _{3}\mathbf{w}_{k:} \\
+\lambda _{1}(\mathbf{L}_{a}\mathbf{U})_{i:}+\lambda _{2}\mathbf{u}_{i:}
\end{multline*}%
%
%V
\begin{equation*}
\nabla _{\mathbf{v}_{i:}}\mathcal{L}=(\mathcal{C}\times _{1}\mathbf{u}%
_{i:}\times _{2}\mathbf{v}_{j:}\times _{3}\mathbf{w}_{k:}-t_{ijk})\mathcal{C}%
\times _{1}\mathbf{u}_{i:}\times _{3}\mathbf{w}_{k:}
\end{equation*}

%W
\begin{equation*}
\nabla_{\mathbf{w}_{k:}}\mathcal{L}=(\mathcal{C} \times_1 \mathbf{u}_{i:}
\times_{2} \mathbf{v}_{j:} \times_{3} \mathbf{w}_{k:}-t_{ijk})\mathcal{C}%
\times_1 \mathbf{u}_{i:} \times_{2} \mathbf{v}_{j:}
\end{equation*}
%CORE 
\begin{equation*}
\nabla _{\mathcal{C}}\mathcal{L}=(\mathcal{C}\times _{1}\mathbf{u}%
_{i:}\times _{2}\mathbf{v}_{j:}\times _{3}\mathbf{w}_{k:}-t_{ijk})\mathbf{u}_{i:}\circ \mathbf{v}_{j:}\circ \mathbf{w}_{k:}
\end{equation*}%
Algorithm \ref{pseudocode} presents the pseudo code. By integrating
the features from different OSNs and exploiting the information of the users
with similar behavior, $\mathbf{U}$ serves as the latent
feature factor of all users from every data source for
semi-supervised learning. 
\begin{algorithm}[t]
\caption{Tensor Factorization for SNMDD}
\label{pseudocode}
\begin{algorithmic}[1]
\renewcommand{\algorithmicrequire}{\textbf{Input:}}
\renewcommand{\algorithmicensure}{\textbf{Output:}}
\REQUIRE Tensors $\mathcal{T}$, an error threshold $\epsilon$, and the max iteration times $I_{Max}$
\ENSURE Low rank matrix $\textbf{U}$, and matrices $\textbf{V}$, $\textbf{W}$, core tensor $\mathcal{C}$

\STATE Initialize $\textbf{U}\in\mathbb{R}^{N\times R}$, $\textbf{V}\in\mathbb{R}^{D\times S}$, $\textbf{W}\in \mathbb{R}^{D \times T}$ , and $\mathcal{C} \in \mathbb{R}^{R\times S \times T}$ with small random values.
  \STATE Set $\eta$ as step size
  \STATE $d_{ii}=\sum_i z_{ij}$
  \STATE $\textbf{L}_z$ = $\textbf{D}- \textbf{Z}$
   \WHILE {$t < I_{Max}$ and $Loss_t - Loss_{t+1} > \epsilon$}
    \FOR {each $t_{ijk}$}
    \STATE Get $\nabla_{u_{i:}}\mathcal{L}$, $\nabla_{v_{j:}}\mathcal{L}$, $\nabla_{w_{k:}}\mathcal{L}$, $\nabla_{\mathcal{C}}\mathcal{L}$
    \STATE $\textbf{u}_{i:}^{t+1}=\textbf{u}_{i:}^{t}-\eta \nabla_{\textbf{u}_{i:}^{t}}\mathcal{L}$
    \STATE $\textbf{v}_{j:}^{t+1}=\textbf{v}_{j:}^{t}-\eta \nabla_{\textbf{v}_{j:}^{t}}\mathcal{L}$
    \STATE $\textbf{w}_{k:}^{t+1}=\textbf{w}_{k:}^{t}-\eta \nabla_{\textbf{w}_{k:}^{t}}\mathcal{L}$
    \STATE $\mathcal{C}^{t+1}=\mathcal{C}^{t}-\eta \nabla_{\mathcal{C}^{t}}\mathcal{L}$
    \ENDFOR
   \ENDWHILE  
\end{algorithmic}
\end{algorithm}
\vspace{-2mm}

In the following, we briefly summarize semi-supervised learning for SNMDD.
Let $\mathbf{x}_{i}$ denote the feature vector of user $i$, and $\mathbf{x}%
_{i}=\mathbf{u}_{i:}$. Let $\mathbf{y}_{i}$ denote the class label vector of
user $i$ with length $K=3$ (i.e., three types of SNMDs), and $\widehat{y}%
_{ik}=1$ indicates that user $i$ suffers from type $k$ of SNMD. For example, 
$\{+1,-1,-1\}$ represents a user with Cyber-Relationship Addiction but
without Net Compulsion and Information Overload. Given a vector set $%
\mathcal{D}$ of $L$ labeled samples $\{\mathbf{x}_{i},\mathbf{y}%
_{i}\}_{i=1}^{L}$ and $L^{\prime }$ unlabeled samples $\{\mathbf{x}_{j},\hat{%
\mathbf{y}}_{j}\}_{j=L+1}^{L+L^{\prime }}$, the optimization problem of TSVM
can be formulated as follows: 

\begin{eqnarray*}
&\smash{\displaystyle\min_{\theta,\xi _{ik}}}&\frac{1}{2}\Vert \mathbf{w}%
\Vert +C\sum_{i=1}^{N}\xi _{ik}+C^{\ast }\sum_{i=L+1}^{L+L^{\prime }}\xi
_{ik}  \notag \\
&\text{s.t. }&\mathbf{y}_{ik}(\mathbf{w}_{k}^{T}\mathbf{x}%
_{i}+b_{k})\geq 1-\xi _{ik}\text{, }1\leq i\leq L,  \notag \\
&&|\hat{\mathbf{y}}_{ik}(\mathbf{w}_{k}^{T}\mathbf{x}_{i}+b_{k})|\geq 1-\xi
_{ik}\text{, }L+1\leq i\leq L+L^{\prime },  \notag \\
&&\xi _{ik}\geq 0\text{, }1\leq i\leq L+L^{\prime }\text{ and }\hat{\mathbf{y%
}}_{ik}\in \{+1,-1\},
\end{eqnarray*}

\noindent where {$\xi _{ik}$} is the slack variable set, and $C$ and $%
C^{\ast }$ are the trade-off parameters between the classification margin
and misclassification errors for labeled samples and unlabeled samples,
respectively. The model parameters $\mathbf{w}_{k}$ and $b_{k}$ returned by
this binary learning problem represent a binary classifier associated with
the $k$-th class: $f_{k}(\mathbf{x}_{i})=\mathbf{w}_{k}^{T}\mathbf{x}%
_{i}+b_{k}$. The binary classifiers for the three SNMD types are trained and used
independently to predict the label vector $\hat{\mathbf{y}}$ for an
unlabeled instance $\mathbf{x}$. The experimental results in the next
section discover that the accuracy of the semi-supervised learning without tensor is $78.3\%$ and $83.1\%$ for Instagram and Facebook, while \textit{STM} increases the accuracy to $89.7\%$ by integrating data from Facebook and Instagram.

\section{Experimental Results}
\label{sec5:exp}
%\linespread{0.96}
In this section, we evaluate SNMDD with real datasets. A user
study with 3126 people is conducted to evaluate the accuracy of SNMDD. Moreover, a feature study is performed. Finally, we apply SNMDD on large-scale datasets and analyze the detected SNMD types.

\subsection{Data Preparation and Evaluation Plan}
\label{DataPrep}
In the following, we detail the preparation of the datasets used in our evaluation.
\subsubsection{User Study}
We recruit 3126 OSN users around the world via Amazon Mechanical Turk (MTurk) to obtain data for training and testing the classifiers in SNMDD. The participants include 1790 males and 1336 females. Their professions are very diverse, affiliating with universities, government offices, technology companies, art centers, banks, and businesses. Each user is first invited to fill out the standard SNMD questionnaires \cite{YoungType98,andreassen12}.\footnote{The IRB number of this project is AS-IRB-HS 15003 v.1.} Then, a group of professional psychiatrists participating in this project assess and manually label the users as \emph{potential SNMD cases} (and identify their types of SNMDs) or \textit{normal users}.\footnote{They are from California School of Professional Psychology, Taipei City Hospital, Nat'l Taipei Univ., psychiatric clinics, etc.} There are 389 users labeled
as \textit{SNMD}, including 246 Cyber-Relationship (CR) Addictions, 267 Information Overload (IO), and 73 Net Compulsion (NC).\footnote{Note that a person may have multiple types of SNMDs simultaneously.} The result obtained by the psychiatrists serves as the ground truth for our evaluation. We also crawl the Facebook (denoted as FB\_US) and Instagram (denoted as IG\_US) data of the participants in the user study for training and testing of our machine learning models (based on features detailed in Section \ref{sec:031_feature}). All the data are collected with the Facebook and Instagram APIs as listed in Table \ref{crawldata}.

In the experiment, we first evaluate the effectiveness of proposed features, including all features (\texttt{All}), social interaction features (\texttt{Social}) and personal profile features (\texttt{Personal}), with a baseline feature \texttt{Duration}, i.e., the total time spent online, using TSVM \cite{TSVM10} for semi-supervised learning in the user study. We also collect two large-scale datasets, including Facebook (denoted as FB\_L) with 63K nodes, 1.5M edges, and 0.84M wall posts \cite{MISLOVE09}, and Instagram (denoted as IG\_L) with 2K users, 9M tags, 1200M likes, and 41M comments \cite{igdata}. Note that some proposed features cannot be extracted from certain large-scale datasets, e.g., game posts and stickers are not available in IG\_L, which is handled by using the imputation technique \cite{MissingData07}. The details of the data crawled from each social media are listed in Table \ref{crawldata}.

\begin{table}[t]
\caption{Details of the datasets}
\label{crawldata}%
\begin{tabular}{p{0.8cm}p{7.4cm}}
\hline
Dataset & Description\\ \hline\hline
FB\_US & User profile, the friends of each user, the news feeds created by users with metadata (who likes, who comments, stickers, and geotag), the news feeds users like or comment (stickers also), events (join/decline), joined groups with events, and game posts created by game apps\\
IG\_US & User profile, the followers/followees of each user, the media created by users with metadata (who likes, who comments, and geotag), and the contents users like or comment\\
FB\_L & Anonymized user ID that performs the action, anonymized user ID that receives the action, and timestamp of action creation\\
IG\_L & Anonymized media ID, anonymized ID of the user who created the media, timestamp of media creation, set of tags assigned to the media, number of likes and number of comments received\\
\hline
\end{tabular}
\vspace{-2mm}
\end{table}

With labeled (IG\_US and FB\_US) and unlabeled data (IG\_L and FB\_L) described above, we divide data into 5 folds with the labeled/unlabeled ratio is preserved. We use 5-fold cross validation, i.e., take 4 folds for training and 1 fold for testing, to evaluate the performance of proposed features using semi-supervised TSVM \cite{5fold}. A number of supervised approaches, including J48 Decision Tree Learning \cite{J48}, SVM \cite{SVMsoft}, and Logistic Regression, and DTSVM \cite{DTSVM10} which do not use unlabeled data, are also compared to justify our choice of using TSVM in SNMDD. Next, we compare the proposed \textit{SNMD-based Tensor Model (STM)} with two baseline algorithms. The first baseline algorithm is to simply concatenate the features from different networks together (denoted as \textit{CF}), while the second baseline algorithm is to use the existing Tucker model (denoted as \textit{Tucker}) which does not incorporate prior knowledge regarding the characteristics of SNMD cases (observed from our analysis). Finally, the effectiveness of each feature is carefully analyzed in Section \ref{featurestudies}.

\subsubsection{Large-Scale Experiments}
To discover new insights, we apply our semi-supervised SNMDD on IG\_L and FB\_L to classify their users and then analyze the detected cases of different SNMD types. Notice that the goal of this analysis is exploratory-oriented as we do not have the ground truth for the large datasets. We examine whether friends of SNMD cases tend to be potential SNMD cases as well. Also, we apply community detection on FB\_L and IG\_L to derive the relationships between different types of SNMD users in their communities. Finally, the average hop distance between the SNMD users of the same type is reported.

\subsection{Evaluation of the Proposed Features}
\label{SNAD_performance}
In the following, we first evaluate the performance of the proposed features using TSVM. We adopt Accuracy (Acc.) and Area Under Curve (AUC) for evaluation of SNMDD. Moreover, Microaveraged-F1 (Micro-F1) and Macroaveraged-F1 (Macro-F1) are also compared for multiple-label classification. Table \ref{InstagramML} summarizes the average results and standard deviations, where the examined feature sets are denoted by self-explained labels.

\begin{table*}[tbp]
\caption{Performance comparisons on the IG\_US and FB\_US datasets.}
\label{InstagramML}%
\centering
%\small
\begin{tabular}{l|llll|llll}
\hline
 & Instagram & & & &  Facebook & & &  \\ \hline\
Measure & \texttt{{Duration}} & \texttt{{Social}} & \texttt{{Personal}} & \texttt{{All}} & \texttt{{Duration}} & \texttt{{Social}} & \texttt{{Personal}} & \texttt{{All}}\\ \hline\hline
Acc. & 0.34$\pm$0.02 & 0.59$\pm$0.01 & 0.69$\pm$0.03 & \textbf{0.78$\pm$0.02} %
     & 0.36$\pm$0.01 & 0.65$\pm$0.05 & 0.73$\pm$0.02 & \textbf{0.83$\pm$0.02} \\
AUC & 0.36$\pm$0.02 & 0.61$\pm$0.01 & 0.74$\pm$0.01 & \textbf{0.79$\pm$0.01} %
         & 0.37$\pm$0.01 & 0.68$\pm$0.01 & 0.77$\pm$0.02 & \textbf{0.84$\pm$0.01} \\
Micro-F1 & 0.42$\pm$0.02 & 0.71$\pm$0.01 & 0.78$\pm$0.04 & \textbf{0.85$\pm$0.01} %
         & 0.44$\pm$0.04 & 0.74$\pm$0.02 & 0.81$\pm$0.01 & \textbf{0.89$\pm$0.01}  \\
Macro-F1 & 0.33$\pm$0.01 & 0.64$\pm$0.01 & 0.73$\pm$0.02 & \textbf{0.85$\pm$0.01} %
         & 0.35$\pm$0.02 & 0.68$\pm$0.02 & 0.77$\pm$0.03 & \textbf{0.90$\pm$0.01} \\
\hline
\end{tabular}%
\vspace{-4mm}
\end{table*}

The results on the IG\_US and FB\_US datasets in the user study show that \texttt{{Duration}} leads to the worst performance, i.e., the results of accuracy are $34\%$ and $36\%$, and the AUC are $0.362$ and $0.379$, respectively. Using all (\texttt{{All}}) or parts (\texttt{{Social}}/ \texttt{{Personal}}) of the features proposed in the paper outperform \texttt{{Duration}} significantly (see Table \ref{InstagramML}). \texttt{{All}} achieves the best performance ($78\%$ and $83\%$ accuracy on the IG\_US and FB\_US datasets, respectively) because SNMDD is able to capture the various features extracted from data logs to effectively detect SNMD cases. Between \texttt{{Social}} and \texttt{{Personal}}, \texttt{{Personal}} outperforms \texttt{{Social}} because the features of temporal behavior (TEMP) in \texttt{{Personal}} are very effective. Since the F1 measure ignores true negatives, its magnitude is mostly determined by the number of true positives, i.e., large classes dominate small classes in microaveraging. As shown in Table \ref{InstagramML}, Micro-F1 of \texttt{{Duration}}, \texttt{{Social}}, and \texttt{{Personal}} are larger than Macro-F1 using both IG\_US and FB\_US datasets, indicating that using parts of features performs better on IO and CR (large classes) than NC. In contrast, the performance of SNMDD is almost the same in Micro-F1 and Macro-F1, which indicates its robustness. The results from FB\_US are better than those from IG\_US because IG\_US is sparser, e.g., there are no event and game posts on Instagram.

After comparing the results from SNMDD with the ground truth obtained via user study, we observe that some false-positive users are detected as NC, probably because people with NC are more likely to hide their real usage time, e.g., the game logs of some people with NC are hidden. As a result, a few normal users may be incorrectly detected as NC. However, SNMDD generally performs very well for NC due to some effective features. For example, users of NC are usually less parasocial since they are less frequent to interact with friends. Moreover, since the NC users' friends with game benefits usually do not know the NC users' other friends (e.g., colleagues), their clustering coefficients are lower than the normal users. %Figure \ref{?} shows the SNMDD results when different percentages of data are labeled. The performance slightly decreases as the number of labeled training samples decreases.

\subsection{Evaluation of Classification Techniques and STM}
In the following, given all the proposed features, we first evaluate TSVM in comparison with some representative  supervised learning approaches in SNMDD. As shown in Table \ref{machinelearningapproach}, the accuracy of semi-supervised TSVM (83.1\%) outperforms all the supervised algorithms, including 76.4\% for $\ell_2$-regularized logistic regression, 77.9\% for $\ell_2$-regularized $\ell_2$-loss SVM, since TSVM effectively uses unlabeled data to address the issues of overfitting and data sparsity. It is worth noting that the accuracy and AUCs of the supervised learning methods, ranging from 74.4\% to 77.9\% and 0.75 to 0.78, respectively, are not significantly different. This is because the proposed features provide robust information so that the accuracy is not sensitive to the choice of learning algorithms.

Next, we compare the proposed \textit{STM} with two baseline algorithms, i.e., \textit{CF} and \textit{Tucker}, to integrate features extracted from both of the IG\_US and FB\_US datasets for learning of classification models using TSVM, as described in Section \ref{DataPrep}. Also as shown in Table \ref{machinelearningapproach}, the accuracy and AUC of \textit{STM} are 89.7\% and 0.926. The results indicate that \textit{STM}, via the decomposed latent factor matrix $\mathbf{U}$, is able to recover some missing features and provide extra latent information to better characterize the users. In contrast, \textit{CF}, which simply concatenates features from FB\_US and IG\_US, performs the worst. Its accuracy and AUC are 75.5\% and 0.759, respectively, even worse than those of some single-source learning algorithms using only the FB\_US dataset, e.g., 77.9\% and 0.783 for $\ell_2$-regularized $\ell_2$-loss SVM. This is because \textit{CF} loses the correlations in some features and thus introduces noises. On the other hand, while \textit{Tucker} is able to achieve 85.6\% accuracy and 0.872 AUC, its performance is still not as great as \textit{STM}. This result indicates that \textit{STM}, by incorporating important prior knowledge about characteristics of SNMD cases (see discussion regarding Eq. (2)), is able to derive more precise and accurate latent features than \textit{Tucker} to achieve the best performance in SNMDD.

\begin{table}[t]
\centering
\caption{Comparisons of SNMDD with different classification techniques.}
\label{machinelearningapproach}%
%\small
\begin{tabular}{lll}
\hline
Technique                                       &  Acc.          &  AUC      \\ \hline\hline
Single-source (FB)                              &                &                \\\hline
J48 Decision Tree Learning                      &  74.4\%        & 0.750          \\
$\ell_1$-regularized $\ell_2$-loss SVM          &  77.6\%        & 0.781          \\
$\ell_2$-regularized $\ell_2$-loss SVM          &  77.9\%        & 0.783          \\
$\ell_1$-regularized logistic regression        &  76.3\%        & 0.776          \\
$\ell_2$-regularized logistic regression        &  76.4\%        & 0.777          \\ 
DTSVM                                           &  76.4\%        &  0.774         \\ 
TSVM                                            &\textbf{83.1\%} & \textbf{0.842} \\ \hline
Multi-source (FB+IG)                            &                &                \\ \hline
CF                                              &  75.5\%        &  0.759         \\
Tucker                                          &  85.6\%        &  0.872         \\
STM                                             &\textbf{89.7\%} & \textbf{0.926} \\\hline
\hline
\end{tabular}%
\vspace*{-4mm}
\end{table}

\subsection{Feature Study}
\label{featurestudies}
To observe the differences among the three types of SNMDs, Table \ref{top5feature} lists the top-5 discriminative features and corresponding accuracy on the FB\_US dataset by TSVM, where CC, BI, BL, and SD respectively denote the clustering coefficient, burst intensity, burst length, and standard deviation. It is worth noting that the number of selfies, an indicator of self-disclosure, is not useful for detecting CR and IO, but it is effective for NC. This is because NC users are usually less socially active as compared to CR and IO users. Moreover, the online/offline interaction ratio of NC is much higher than the ratios of the other two types, probably because NC users usually show less willingness to join offline activities. In contrast, users of CR and IO prefer to use social media instead of playing games alone. Moreover, people with compulsive personality are more introverted. In contrast, people with CR usually create virtual bonds to develop pathological relationships for compensation of their (missing) offline relationship. 

The parasociality, effective for detecting all SNMD types, is especially useful for detecting CR cases. For example, in our user study, we find user A, 21-year-old male, frequently posting news feeds, such as ``I'm so bored :(((((...Ahhhhhh!!'', and his cross-dressing photos on his Facebook timeline, more than 3 times a week, which usually get less than 5 likes. At the same time, he ``likes'' a large number of posts from others. SNMDD classifies him as a potential CR case and his questionnaire reveals that he constantly blocks out disturbing thoughts about life and finds himself anticipating when he goes online again.

Burst intensity and length seem to be quite useful for detecting IO cases. For example, user B, 36-year-old male, is detected as IO since the behavior of clicking ``likes'' fits the pattern of bursts, i.e., the median of his burst intensity is high, equal to 31. His answers to the standard questionnaire reveal that he loses sleep due to late-night access on Facebook to check others' news feeds. Through interview, user B explains that he cannot stop checking for new posts and e-mails even when all his news feeds and emails are read. Some of his friends reply him: ``are you a robot? no sleep needed?!?!!'', indicating that user B is indulged in finding social news.

Next, we analyze the importance of different features to our classifiers. The information gain is exploited to measure the importance of each feature. In summary, the top 5 important
features overall are : 1) median of the intensity of bursts, 2)
online/offline interaction ratio, 3) parasociality, 4) number of used
stickers, and 5) standard deviation of the length of bursts. It is worth noting
that TSVM using only these 5 features in SNMDD achieves an
accuracy of 76.4\% and 80.7\% for IG\_US and FB\_US, respectively, close to that
of using all features (\texttt{{All}}). In other words, integrating important social and personal features provides good results because effective personal features, e.g., the temporal behavior features, can be used to differentiate the users suffering from withdraw or relapse symptoms and heavy users, while social features capture the interactions among users to differentiate different SNMDs.

\begin{table}[t]
\caption{Top features and Acc. on the FB\_US dataset.}
\label{top5feature}%
\small
\begin{tabular}{lll}
\hline
CR & NC & IO \\ \hline\hline
Parasociality                   & Game posts      & Median of BI\\
Median of BI     &   Online/offline ratio    & Online/offline ratio \\
Sticker number     & Parasociality     & SD of BL\\
Online/offline ratio    & Number of selfies    & Sticker number\\
CC.    & CC.    & Parasociality\\
Acc.: 80.2\% & Acc.: 76.8\% & Acc.: 82.7\% \\ \hline
\end{tabular}%
\vspace{-2mm}
\end{table}

Finally, we carefully examine the effectiveness of each feature on the FB\_US dataset. Table \ref{featureeffectiveness} compares the performance of different feature combinations using TSVM, where \texttt{{All}}--$X$ means all features \emph{excluding} category $X$. Unsurprisingly, combining all features leads to the best performance with the accuracy of $83.1\%$. In terms of individual category of features, the temporal behavior features are the most effective, whereas the profile features are the least effective. The accuracy obtained by excluding one category of features is at least $68.1\%$, which shows that the features are generally robust even when one feature set is missing. The accuracy of \texttt{{All}}--PROF is close to \texttt{{All}}, indicating that PROF is the least important.

\begin{table}[t]
\centering
\caption{Feature effectiveness analysis: SNMDD accuracy on the FB\_US dataset.}
\label{featureeffectiveness}%
\small
\begin{tabular}{llll}
\hline
Used Features & Accuracy & Used Features & Accuracy   \\ \hline\hline
PR       &    56.9\%         & \texttt{{All}}--PR       & 78.2\%  \\ %
ONOFF    &    60.3\%         & \texttt{{All}}--ONOFF    & 75.1\%  \\ %
SC       &    40.1\%         & \texttt{{All}}--SC       & 78.8\%  \\
SSB      &    44.4\%         & \texttt{{All}}--SSB      & 79.3\%  \\
SD       &    58.9\%         & \texttt{{All}}--SD       & 73.2\%  \\ %
TEMP     &    67.5\%         & \texttt{{All}}--TEMP     & 68.1\%  \\ % 
UT       &    36.4\%         & \texttt{{All}}--UT       & 82.6\%  \\
DIS      &    54.0\%         & \texttt{{All}}--DIS      & 75.9\%  \\
PROF     &    18.2\%         & \texttt{{All}}--PROF     & 81.5\%  \\
\texttt{{All}}      &    83.1\%         &              &         \\ \hline
\end{tabular}%
\vspace{-2mm}
\end{table}

\begin{figure}[t]
\centering
\subfigure[][Relative improvement w.r.t. the number of features.] {\  \includegraphics[scale=0.141] {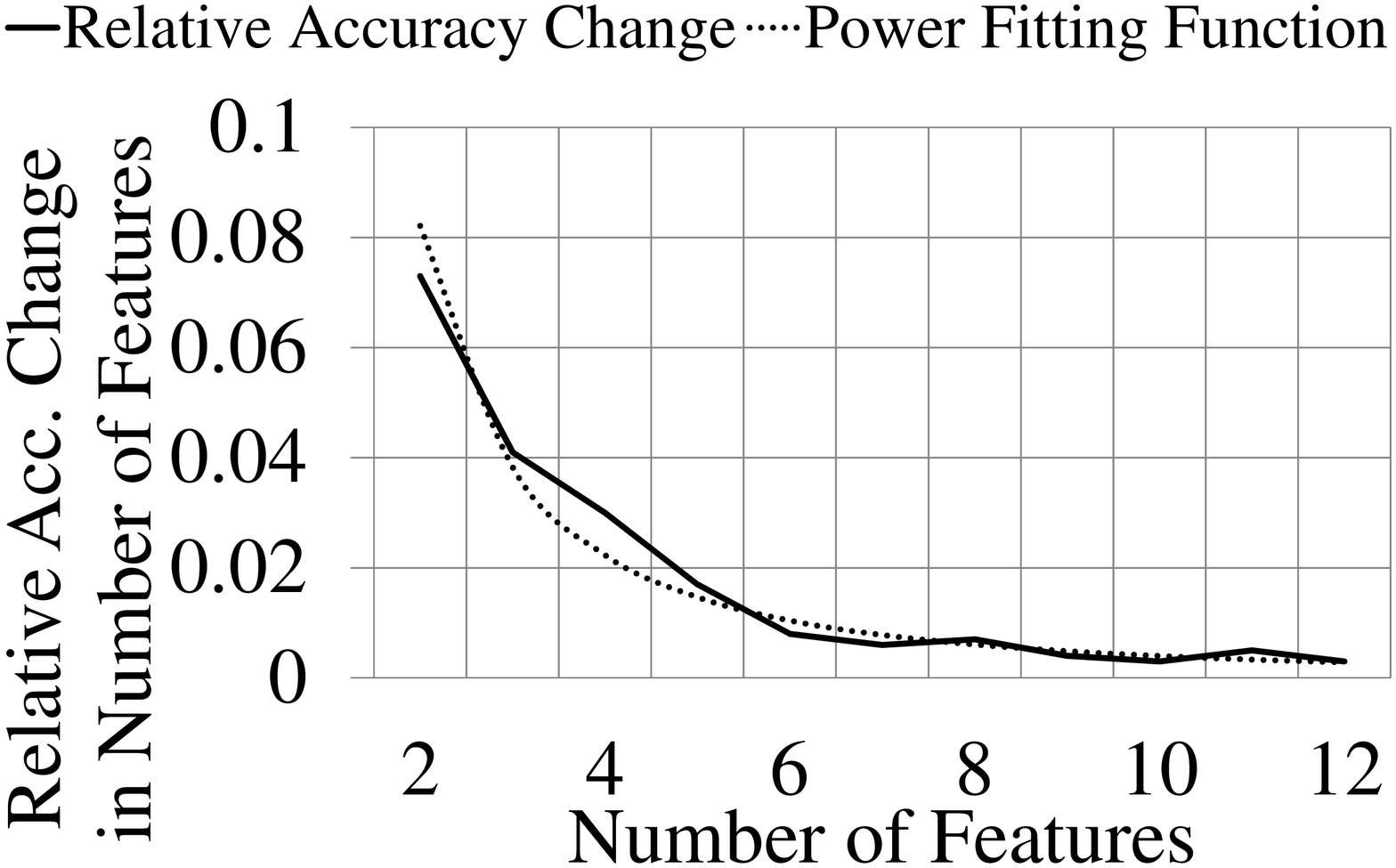}
\label{DR} } 
\subfigure[][Relative improvement w.r.t. the number of latent features (\textit{STM}).] {\  \includegraphics[scale=0.141] {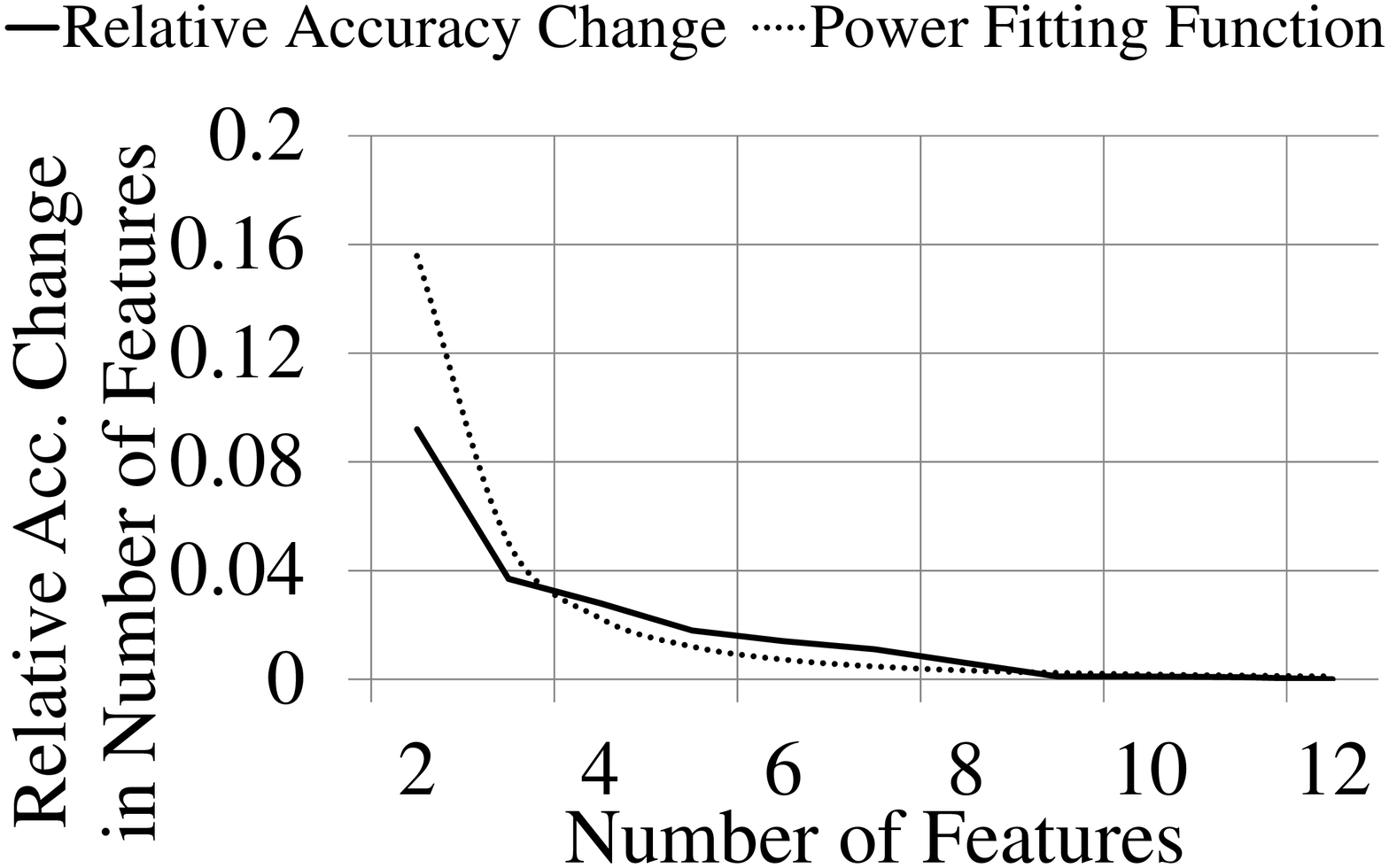}
\label{DR2} }
\vspace{-1mm}
\caption{Relative accuracy change with respect to number of features.}
\vspace{-1mm}
\end{figure}

Figs. \ref{DR} and \ref{DR2} show the improvement of adding different features in TSVM on the FB\_US dataset and the proposed \textit{STM} on multi-source data (i.e., FB\_US and IG\_US). The feature selection of TSVM is based on the information gain (the top-5 features mentioned earlier), while the tensor approach automatically extracts important latent features. We observe a diminishing return property on both figures, where the improvement becomes marginal as more features are included. Fig. \ref{DR} shows a power fit function ($p(x) = 0.3087x^{-1.89}$) of the curve with $R^2 = 0.9512$. The exponent $-1.89$ denotes that the improvement by adding $n$-th feature is $n^{-1.89}$ times smaller than that by adding the first feature. On the other hand, the results of the tensor-based approach in Fig. \ref{DR2} show that the accuracy increment for adding a single feature drops faster ($p(x) = 1.06x^{-2.82}$) since the proposed \textit{STM} can extract much more important and concise features.

\subsection{Analysis of SNMD Types in Large Datasets}
\label{Sec:SNAlargeSNMDD}

\begin{figure}[t]
\centering
\subfigure[][SNMD types of friends (FB\_L).] {\  \includegraphics[scale=0.15] {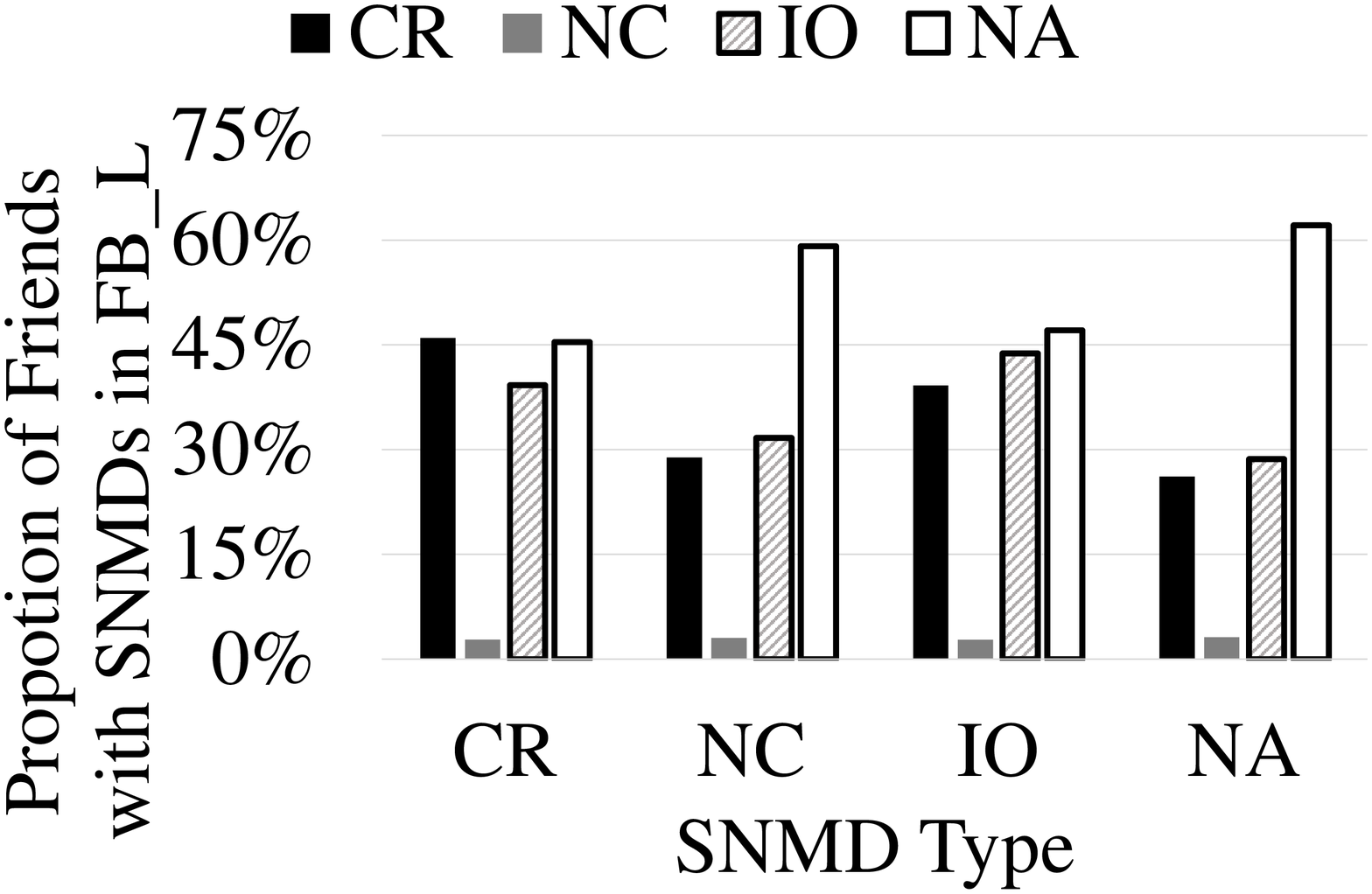}
\label{fb_typefriend} } 
\subfigure[][SNMD types of friends (IG\_L).] {\  \includegraphics[scale=0.15] {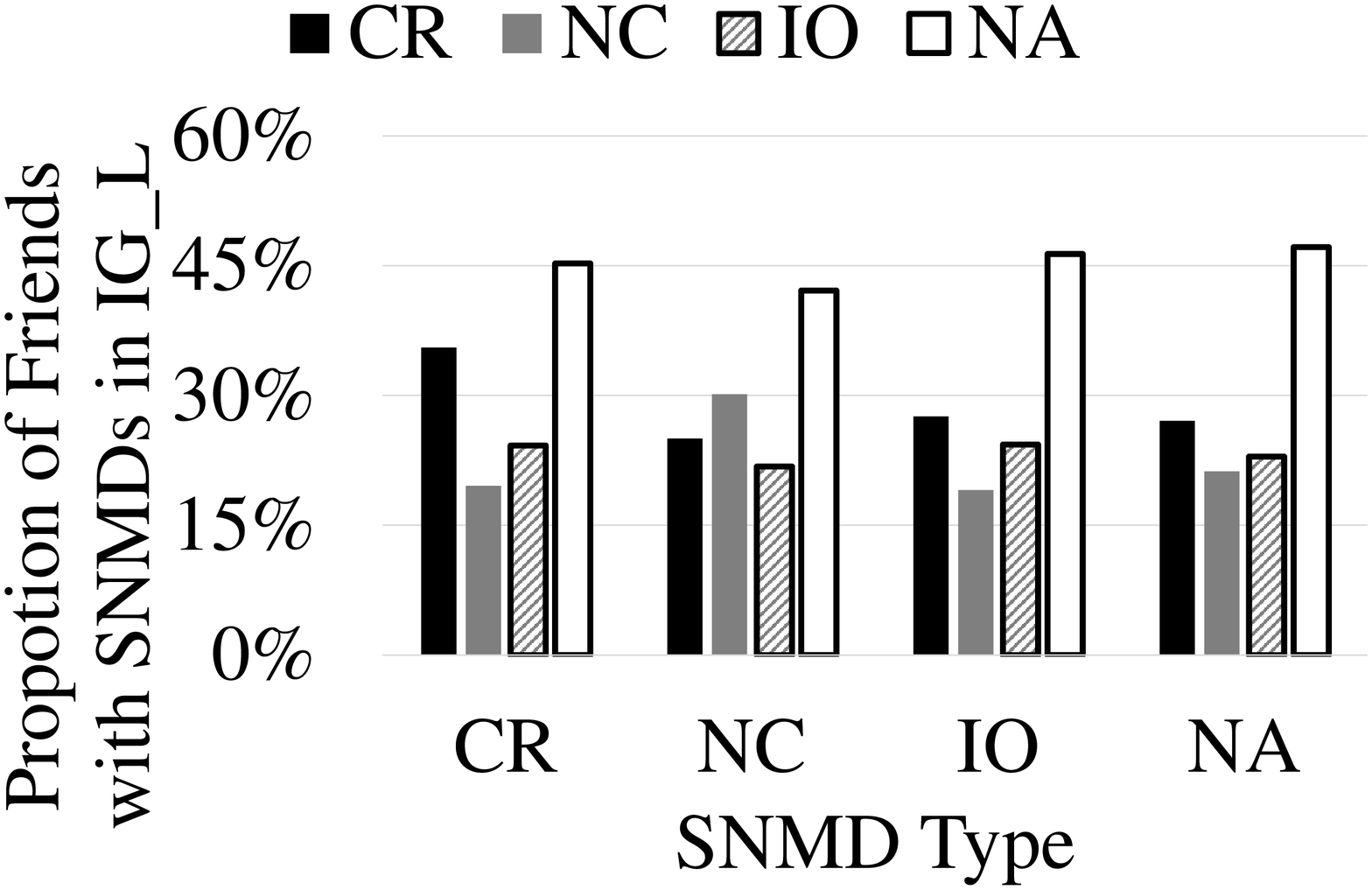}
\label{ig_typefriend} } 
\subfigure[][Ratios of SNMD users in communities (FB\_L).] {\  \includegraphics[scale=0.15] {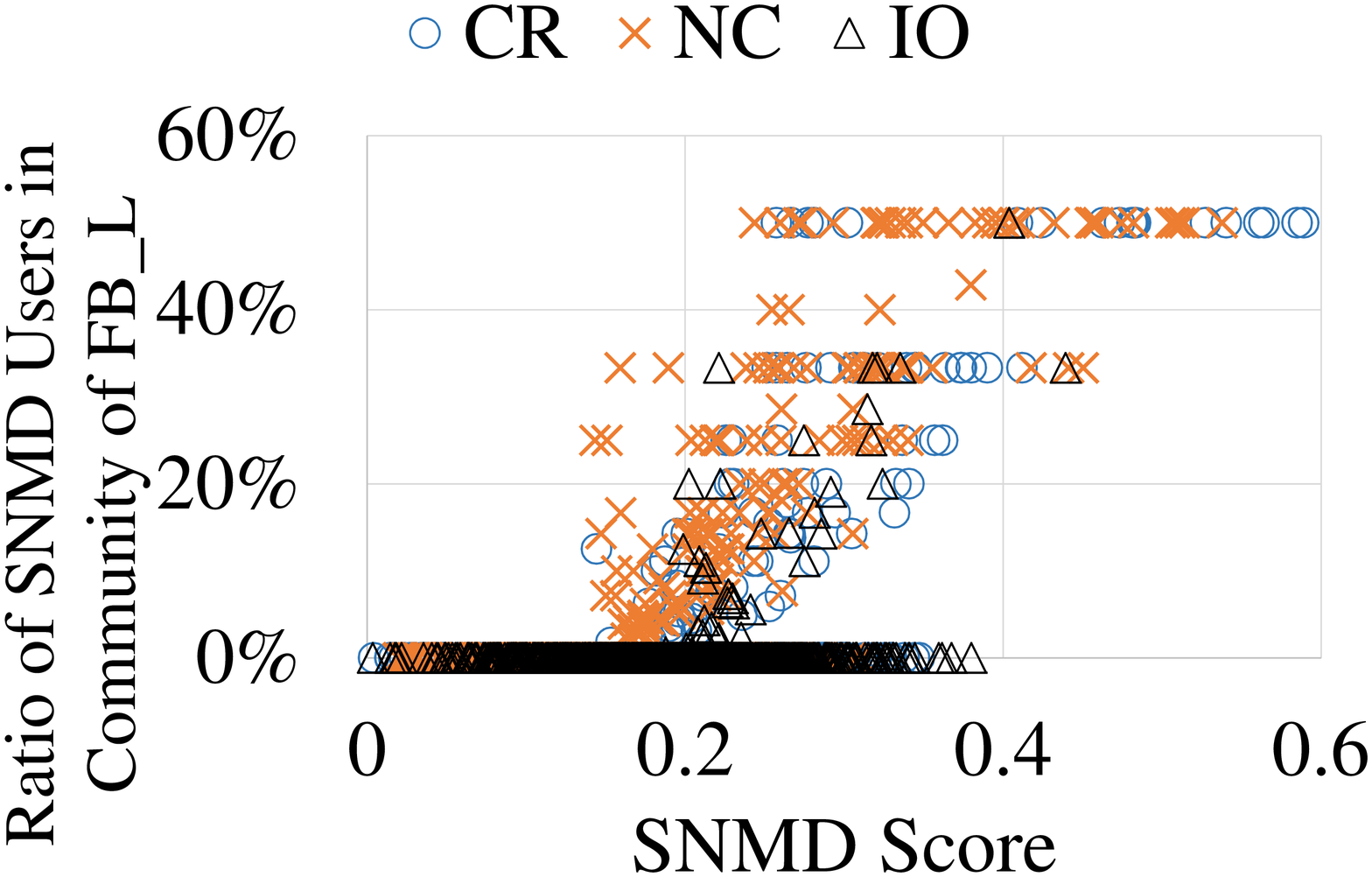}
\label{fb_ratiocomm} } 
\subfigure[][Ratios of SNMD users in communities (IG\_L).] {\  \includegraphics[scale=0.15] {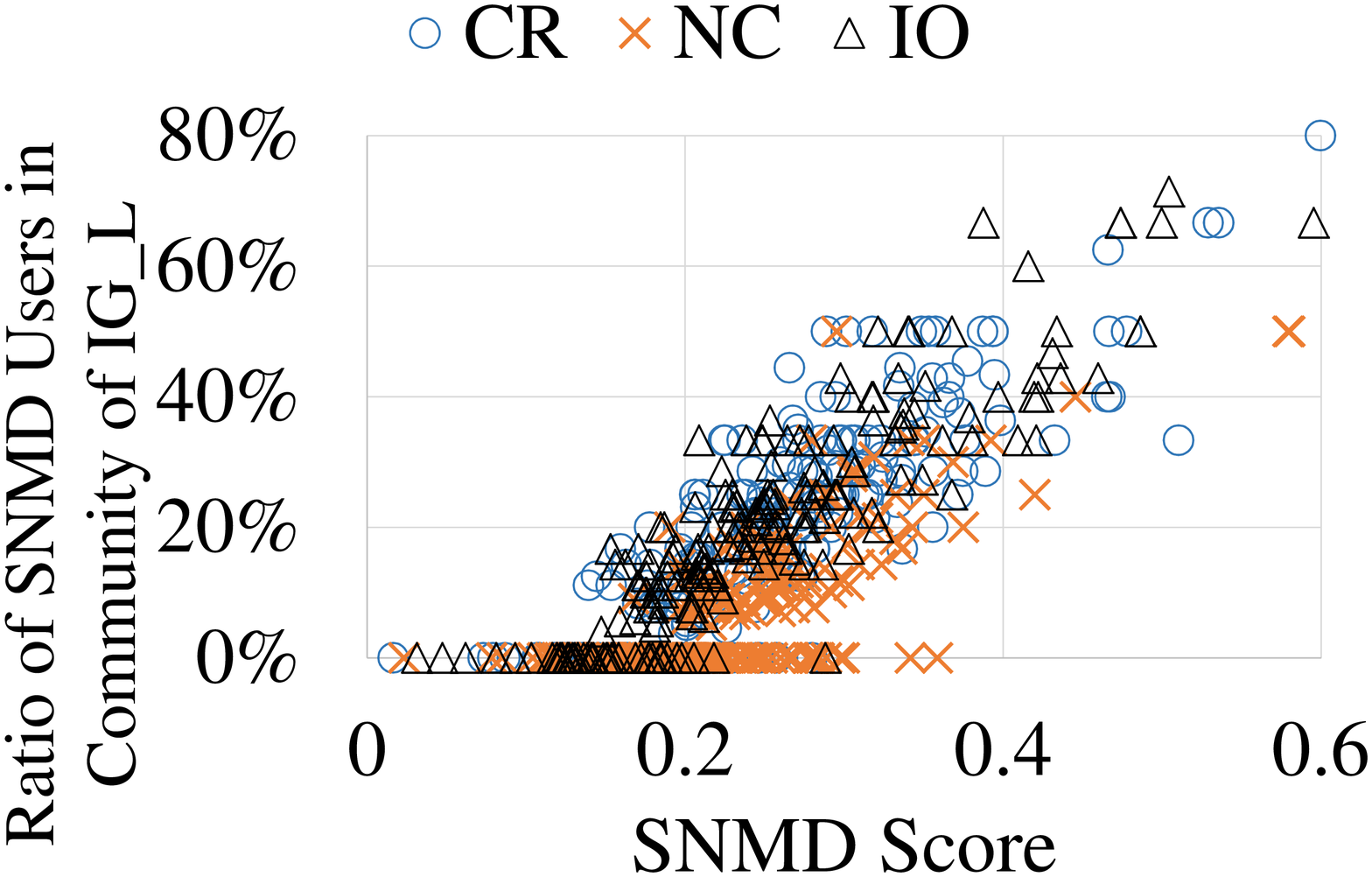}
\label{ig_ratiocomm} } 
\subfigure[][SNMD users.] {\  \includegraphics[scale=0.15] {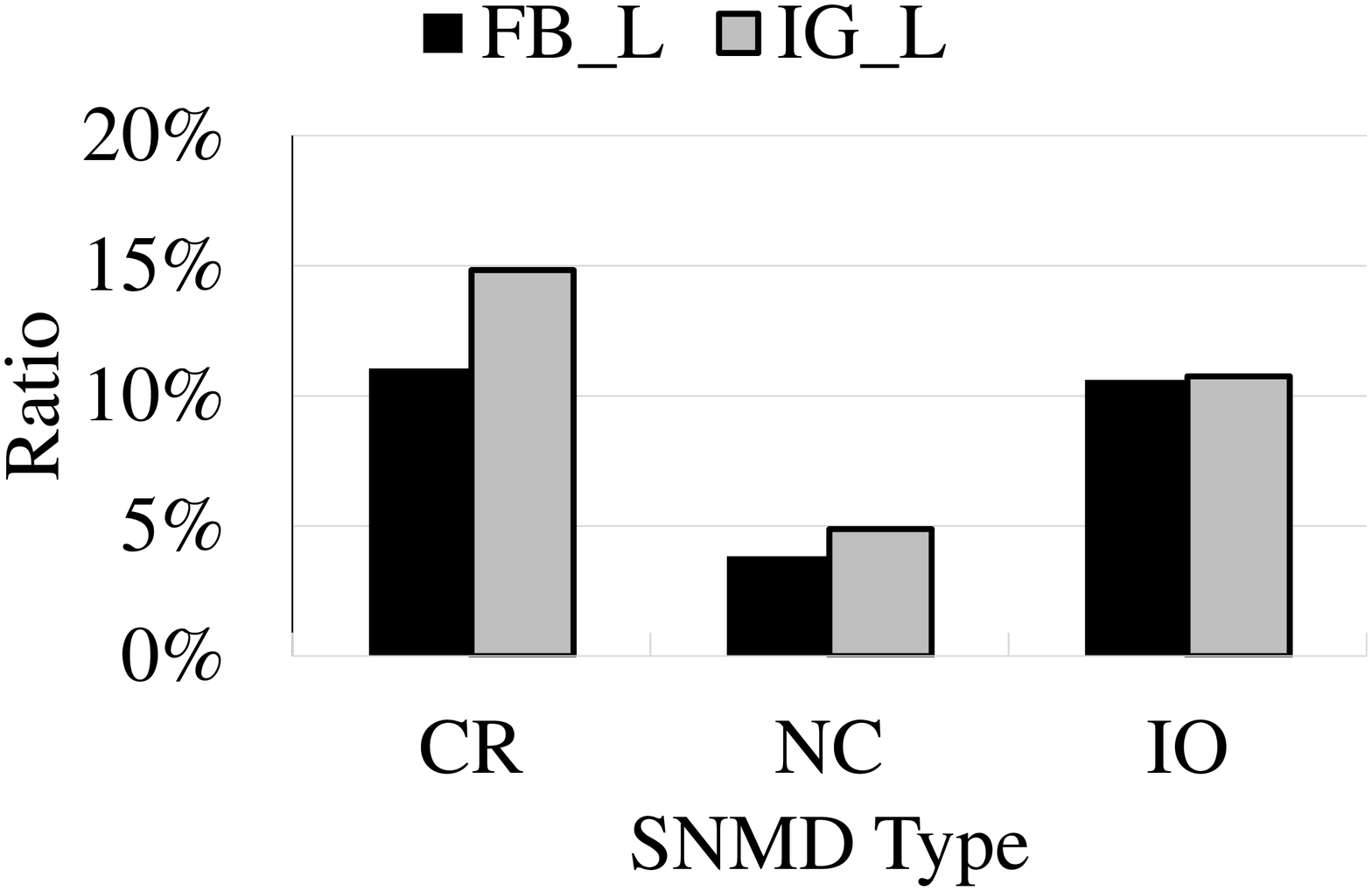}
\label{typeconst} } 
\subfigure[][Hop distance to same type SNMD users.] {\  \includegraphics[scale=0.15] {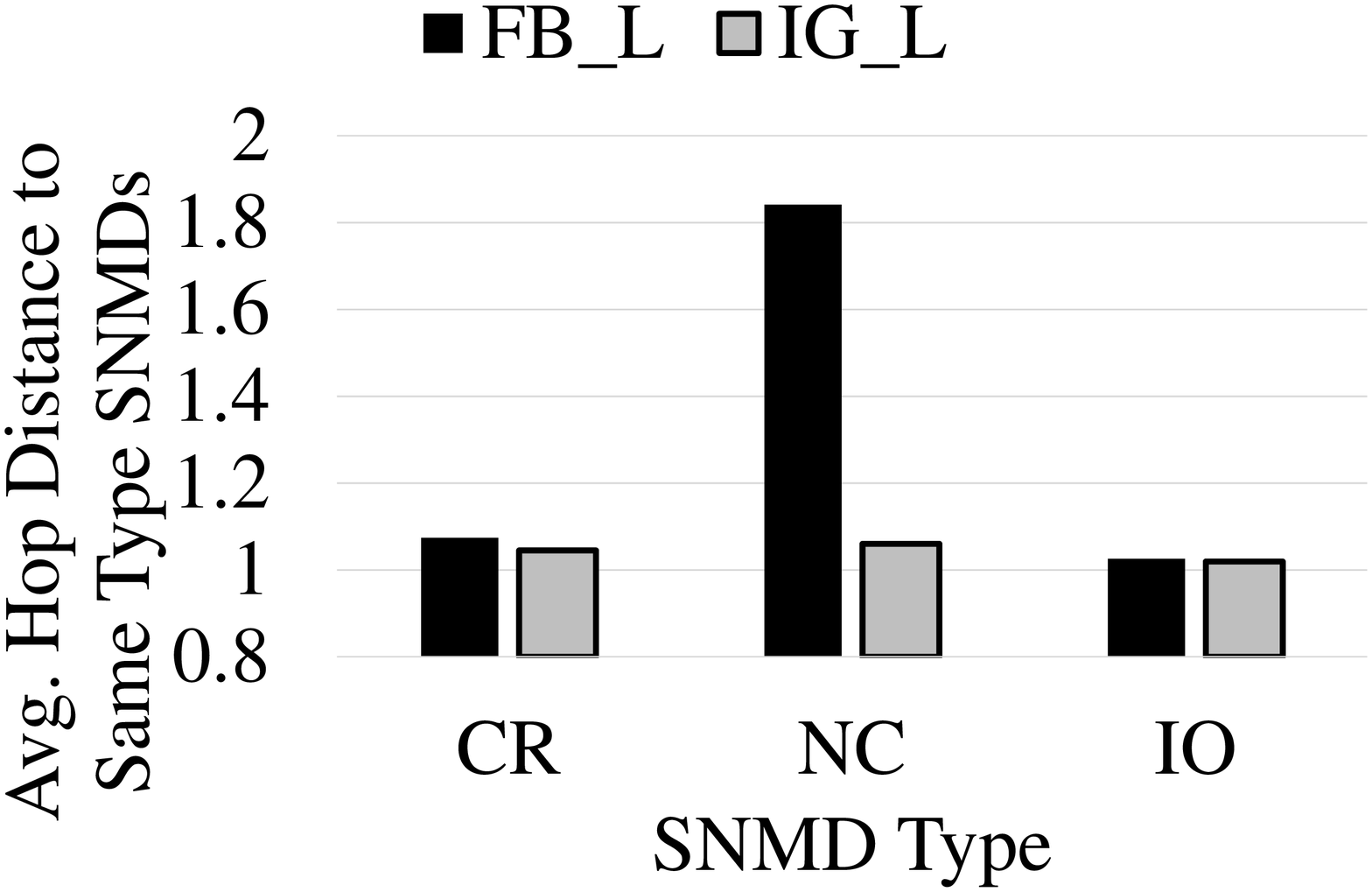}
\label{hopdist} }
\vspace{-2mm}
\caption{Comparisons of different datasets.}
\label{exp_opt}
\vspace{-3mm}
\end{figure}

%\begin{figure}
%\centering
%\includegraphics[width=3.5cm,height=2cm] {FIG_visual.eps}
%\caption{Visualization of NC users.}
%\label{visual}
%\end{figure}
In this analysis, we first apply the proposed SNMDD framework (with TSVM) on some large-scale OSN datasets, i.e., FB\_L and IG\_L, to classify their users. In Figs. \ref{fb_typefriend} and \ref{ig_typefriend}, we analyze the detected SNMD cases among the friends of an SNMD user. In Fig. \ref{fb_typefriend}, the leftmost bar indicates that in FB\_L, among all CR users, about 45\% of their friends are also CR users, which is greater than the percentage of other SNMD types. On the other hand, the 8th bar from the left in Fig. \ref{fb_typefriend} indicates that in FB\_L, about 59\% of NC users' friends are NA (non-SNMD users). Figs. \ref{fb_typefriend} and \ref{ig_typefriend} show that, in FB\_L and IG\_L, CR and IO users have similar friend types. This is because CR and IO cases, by their nature, are similar, i.e., they are both seeking social satisfaction (e.g., relationships and information) from the OSNs. Moreover, among different SNMD cases, CR and IO users are likely to be friends with other CR and IO users. For CR users, this phenomenon has been described as ``loneliness propagates'' \cite{Cacioppo09}.

%Furthermore, comparing the results in FB\_L and IG\_L (Figs. \ref{fb_typefriend} and \ref{ig_typefriend}), we find that in IG\_L, NC users usually have a more balanced ratio of different types of SNMD friends. This is because when the NC users use Instagram, they behave like less-active users since there are no gambling or game on Instagram, i.e., they show NC behavior on other OSNs with gaming or gambling. We observe that these NC users in IG\_L are usually identified by SNMDD through their parasocial behavior, i.e., more people comment on their posts, but the NC users less frequently comment on the others'. 

Furthermore, Infomap community detection \cite{INFOMAP08} is performed on FB\_L and IG\_L to derive the relationships between different types of SNMD users in their communities. Figs. \ref{fb_ratiocomm} and \ref{ig_ratiocomm} analyze the community structures of SNMD users with different SNMD scores, where each point represents the characteristic of a community. Specifically, each community in the dataset is represented by three different types of points, i.e., CR, NC, and IO. For example, each CR point is represented as $\langle score, ratio \rangle$, where $score$ is the average CR score in that community, and $ratio$ indicates the proportion of CR users in the community. It is similar for each IO/NC point. As Figs. \ref{fb_ratiocomm} and \ref{ig_ratiocomm} show, for each SNMD type, when the average SNMD score is higher, it is likely to have more SNMD users in the community. Moreover, there are many communities with large IO scores in IG\_L that have IO ratios close to 1. This implies that the users with large IO scores in IG\_L are more inclined to form homogeneous groups. At the first glance, one may feel that NC users frequently appear in many communities, and there seems to be a large number of NC users, especially in FB\_L (i.e., Fig. \ref{fb_ratiocomm}). However, after carefully examining these communities, we find that those communities (with large ratios of NC users) are usually very small (usually with the size around 5) because NC users are less-active. On the other hand, in IG\_L, when SNMD scores are larger, the ratios of IO users in communities are also larger. This is because IO users can view, like, or follow others in Instagram more easily (not necessary to be friends first).

Fig. \ref{typeconst} compares the ratios of different types of SNMD users identified in FB\_L and IG\_L. There are more CR users in IG\_L probably because CR users seek social supports online to compensate the loneliness in real life. We argue that the Instagram platform makes it easy to freely create social relationships with strangers. In contrast, it is not that easy to create new social relationships on Facebook since the friend requests need to be approved. Finally, Fig. \ref{hopdist} compares the average number of hops from each SNMD user to the nearest user with the same type of SNMDs. The leftmost bar shows that the average hop distance from each CR user to the closest CR user is 1.07 hop, indicating that CR and IO users are close to other same-type users, i.e., average hop distances are within 1.15, where Figs. \ref{fb_typefriend} and \ref{ig_typefriend} also report similar results. 

%However, as indicated in Fig. \ref{fb_typefriend}, NC users in FB\_L usually do not have other NC users nearby because NC users focus on games and may not be interested in building friendships.

%Therefore, Fig. \ref{hopdist} shows that the average hop distance between NC users in Facebook is about 2, much larger than the other types. Moreover, Fig. \ref{visual} visualizes several small communities in FB\_L and highlights NC users, where magenta and purple points represent NC users and the others, respectively. Since the NC users in FB\_L usually have a small number of friends in their communities, the average hop distance between users is large.
\vspace{-2mm}
\section{Conclusion}
\label{conclu}
In this paper, we make an attempt to automatically identify potential online users with SNMDs. We propose an SNMDD framework that explores various features from data logs of an OSN and a new tensor technique for deriving latent features from multiple OSNs for SNMD detection. This work represents a collaborative effort between computer scientists and mental healthcare researchers to address emerging issues in SNMDs. As for the next step, we plan to study the features extracted from multimedia contents by techniques on NLP and computer vision. We also plan to further explore new issues from the perspective of a social network service provider, e.g., Facebook or Instagram, to improve the well-beings of OSN users without compromising user engagement.

\section{Acknowledgments}
This work is supported in part by NSF through grants III-1526499, CNS-1115234, and OISE-1129076, USA, and by Ministry of Science and Technology through grants MOST 104-2221-E-002-214-MY3, 103-2221-E-001-005-MY2, and 104-2221-E-001-005-MY2, Taiwan.


\begin{thebibliography}{99}
\bibitem{YoungType98} K. Young, M. Pistner, J. O'Mara, and J. Buchanan. Cyber-disorders: The mental health concern for the new millennium. \textit{Cyberpsychol. Behav.}, 1999.

\bibitem{b08} J. Block. Issues of DSM-V: internet addiction. \textit{American Journal of Psychiatry}, 2008.

\bibitem{Kimberly98} K. Young. Internet addiction: the emergence of a new clinical disorder, \textit{Cyberpsychol. Behav.}, 1998.

\bibitem{Lin14} I.-H. Lin, C.-H. Ko, Y.-P. Chang, T.-L. Liu, P.-W. Wang, H.-C. Lin, M.-F. Huang, Y.-C. Yeh, W.-J. Chou, and C.-F. Yen. The association between suicidality and Internet addiction and activities in Taiwanese adolescents. \textit{Compr. Psychiat.}, 2014.

\bibitem{Baek13} Y. Baek, Y. Bae, and H. Jang. Social and parasocial relationships on social network sites and their differential relationships with users' psychological well-being. \textit{Cyberpsychol. Behav. Soc. Netw.}, 2013.

\bibitem{Barbera09} D. La Barbera, F. La Paglia, and R. Valsavoia. Social network and addiction. \textit{Cyberpsychol. Behav.}, 2009.

\bibitem{Chak04} K. Chak and L. Leung. Shyness and locus of control as predictors of internet addiction and internet use. \textit{Cyberpsychol. Behav.}, 2004.

\bibitem{cacm12} O. Turel and A. Serenko. Is mobile email addiction overlooked? \textit{CACM}, 2010.

\bibitem{Baumer13} E. Baumer, P. Adams, V. Khovanskaya, T. Liao, M. Smith, V. Sosik, and K. Williams. Limiting, leaving, and (re)lapsing: an exploration of Facebook non-use practices and experiences. \textit{CHI}, 2013.

\bibitem{Dinakar11} K. Dinakar, B. Jones, C. Havasi, H. Lieberman, and R. Picard. Common sense reasoning for detection, prevention, and mitigation of cyberbullying. \textit{ACM TiiS}, 2012.

\bibitem{SVM1} R. Jain and N. Abouzakhar. A comparative study of hidden markov model and support vector machine in anomaly intrusion detection. \textit{JITST}, 2013.

\bibitem{SVM3} C. Tan, L. Lee, J. Tang, L. Jiang, M. Zhou, and P. Li. User-level sentiment analysis incorporating social networks. \textit{KDD}, 2011.

\bibitem{Crammer08} K. Crammer, M. Kearns, and J. Wortman. Learning from multiple sources. \textit{JMLR}, 2008.

\bibitem{TSVM10} R. Collobert, F. Sinz, J. Weston, and L. Bottou. Large scale transductive svms. \textit{JMLR}, 2006.

\bibitem{Leung04} L. Leung. Net-generation attributes and seductive properties of the internet as predictors of online activities and internet addiction. \textit{Cyberpsychol. Behav. Soc. Netw.}, 2004.

\bibitem{Cacioppo09} J. Cacioppo, J. Fowler, and N. Christakis. Alone in the crowd: the structure and spread of loneliness in a large social network. \textit{J. Pers. Soc. Psychol.}, 2009.

\bibitem{CheckinD} R. Schwartz. Mobility and locative media: mobile communication in hybrid spaces. \textit{Routledge}, 2014.

\bibitem{Steinfield12} C. Steinfield, N. Ellison, C. Lampe, and J. Vitak. Online social network sites and the concept of social capital. \textit{Frontiers in new media research, New York: Routledge}, 2012.

\bibitem{socialcap01} R. Davis. A cognitive-behavioral model of pathological internet use. \textit{Comput. Hum. Behav.}, 2001.

\bibitem{Arnaboldi13} V. Arnaboldi, M. Conti, A. Passarella, and R. Dunbar. Dynamics of personal social relationships in online social networks: a study on twitter. \textit{ACM COSN}, 2013.

\bibitem{Wise10} K. Wise, S. Alhabash, and H. Park. Emotional responses during social information seeking on Facebook. \textit{Cyberpsychol. Behav. Soc. Netw.}, 2010.

\bibitem{Tamir12} D. Tamir and J. Mitchell. Disclosing information about the self is intrinsically rewarding. \textit{Natl. Acad. Sci.}, 2012.

\bibitem{LiuEmoticon12} K.-L. Liu, W.-J. Li, and M. Guo. Emoticon smoothed language models for twitter sentiment analysis. \textit{AAAI}, 2012.

\bibitem{Seidman13} G. Seidman. Self-presentation and belonging on Facebook: How personality influences social media use and motivations. \textit{Pers. Indiv. Differ.}, 2013.

\bibitem{Kleinberg02} J. Kleinberg. Bursty and Hierarchical Structure in Streams. \textit{KDD}, 2002.

\bibitem{Lai13} C.-M. Lai, K.-K. Mak, H. Watanabe, R. Ang, J. Pang, and R. Ho. Psychometric properties of the internet addiction test in Chinese adolescents. \textit{J. Pediatr. Psychol.}, 2013.	

\bibitem{Klinkner08} R. Jones and K. Klinkner. Beyond the session timeout: automatic hierarchical segmentation of search topics in query logs. \textit{CIKM}, 2008.

\bibitem{Suler04} J. Suler. The online disinhibition effect. \textit{CyberPsychol. Behav.}, 2004.

\bibitem{Wang14} G. Wang, B. Wang, T. Wang, A. Nika, H. Zheng, and B. Y. Zhao. Whispers in the dark: analysis of an anonymous social network. \textit{IMC}, 2014.

\bibitem{Makashvili13} M. Makashvili, B. Ujmajuridze, T. Amirejibi, B. Kotetishvili, and S. Barbakadze. Gender difference in the motives for the use of Facebook. \textit{TOJET}, 2013.

\bibitem{TuckerModel09} T. Kolda and B. Bader. Tensor decompositions and applications. \textit{SIAM review}, 2009.

\bibitem{Chaoji} M. Roth, A. Ben-David, D. Deutscher, G. Flysher, I. Horn, A. Leichtberg, N. Leiser, Y. Matias, and R. Merom. Suggesting friends using the implicit social graph. \textit{KDD}, 2010.

\bibitem{andreassen12} C. Andreassen, T. Torsheim, G. Brunborg, and S. Pallesen. Development of a Facebook addiction scale. \textit{Psychol. Rep.}, 2012.

\bibitem{MISLOVE09} B. Viswanath, A. Mislove, M. Cha, and K. P. Gummadi. On the evolution of user interaction in Facebook. \textit{WOSN}, 2009.

\bibitem{igdata} E. Ferrara, R. Interdonato, and A. Tagarelli. Online popularity and topical interests through the lens of Instagram. \textit{HT}, 2014.

\bibitem{MissingData07} M. Saar-Tsechansky and F. Provost. Handling Missing Values when Applying Classification Models. \textit{JMLR}, 2007.

\bibitem{5fold} F. Shang, L. C. Jiao, and F. Wang. Semi-Supervised Learning with Mixed Knowledge Information. \textit{KDD}, 2012.

\bibitem{J48} I. Witten and E. Frank. Data mining: practical machine learning tools and techniques with Java implementations. Morgan-Kaufmann, San Francisco, 2000.

\bibitem{SVMsoft} C.-C. Chang and C.-J. Lin. LIBSVM: a library for support vector machines, 2001.

\bibitem{DTSVM10} F. Chang, C.-Y. Guo, X.-R. Lin, and C.-J. Lu. Tree decomposition for large-scale SVM problems. \textit{JLMR}, 2010.

\bibitem{INFOMAP08} M. Rosvall and C. Bergstrom. Maps of random walks on complex networks reveal community structure. \textit{Natl. Acad. Sci.}, 2008.

\end{thebibliography}
\end{document}